\documentclass[fleqn,usenatbib]{mnras}

\usepackage{newtxtext,newtxmath}
\usepackage[T1]{fontenc}
\usepackage{ae,aecompl}

\usepackage{graphicx}	% Including figure files
\usepackage{amsmath}	% Advanced maths commands

\usepackage{amssymb}	% Extra maths symbols
\usepackage{array}
\usepackage{booktabs}
\usepackage{float} 
\usepackage{tabularx}
\usepackage{soul,xcolor}

\title[Galaxy cluster alignment]{The spatial distribution of satellites in galaxy clusters}

\author[Gu et al.]{
Qing Gu,$^{1,2}$
Qi Guo,$^{1,2}$\thanks{E-mail: guoqi@bao.ac.cn}
Tianchi Zhang,$^{1,3}$ Marius Cautun,$^{4,5}$ Cedric Lacey,$^{5}$\newauthor
Carlos S. Frenk$^{5}$ and Shi Shao$^{1,5}$\\
$^{1}$Key Laboratory for Computational Astrophysics, National Astronomical Observatories, Chinese Academy of Sciences, Beijing 100101, China\\
$^{2}$School of Astronomy and Space Science, University of Chinese Academy of Sciences, Beijing 100049, China\\
$^{3}$Beijing Planetarium, Beijing Academy of Science and Technology, Beijing 100044, China\\
$^{4}$Leiden Observatory, Leiden University, PO Box 9513, NL-2300 RA Leiden, the Netherlands\\
$^{5}$Institute for Computational Cosmology, Department of Physics, Durham University, South Road, Durham DH1 3LE, UK\\
}

\date{Accepted XXX. Received YYY; in original form ZZZ}

\pubyear{2018}

\begin{document}
\label{firstpage}
\pagerange{\pageref{firstpage}--\pageref{lastpage}}
\maketitle

\begin{abstract}
The planar distributions of satellite galaxies around the Milky Way and Andromeda have been extensively studied as potential challenges to the standard cosmological model. Using the Sloan Digital Sky Survey and the Millennium simulation we extend such studies to the satellite galaxies of massive galaxy clusters. We find that both observations and simulations of galaxy clusters show an excess of anisotropic satellite distributions. On average, satellites in clusters have a higher degree of anisotropy than their counterparts in Milky-Way-mass hosts once we account for the difference in their radial distributions. The normal vector of the plane of satellites is strongly aligned with the host halo's minor axis, while the alignment with the large-scale structure is weak. At fixed cluster mass, the degree of anisotropy is higher at higher redshift. This reflects the highly anisotropic nature of satellites accretion points, a feature that is partly erased by the subsequent orbital evolution of the satellites. We also find that satellite galaxies are mostly accreted singly so group accretion is not the explanation for the high flattening of the planes of satellites. 
\end{abstract}


%%%%%%%%%%%%%%%%%%%%%%%%%%%%%%%%%%%%%%%%%%%%%%%%%%

%%%%%%%%%%%%%%%%% BODY OF PAPER %%%%%%%%%%%%%%%%%%

\section{Introduction}
\label{sec:introduction}
The $\Lambda$ cold dark matter ($\Lambda$CDM) cosmological model has been very successful at reproducing many large-scale observations, such as the power spectrum of the Cosmic Microwave Background radiation \citep[e.g.][]{Hinshaw2013,Planck2016}, the large-scale galaxy clustering  \citep[e.g.][]{Colless2001,York2000,Springel2005,Alam2017} and the accelerated expansion of the Universe (see \citealt{Weinberg2013}). However, challenges persist on small scales, including the missing satellites problem \citep[e.g.][]{Klypin1999b,Moore1999}, the cusp-core problem \citep[e.g.][]{Flores1994,Moore1994} and the too-big-to-fail problem \citep[e.g.][]{Boylan2011}. They, however, can be readily accounted for by a variety of processes involving the baryonic component of galaxies (\citealt{Sawala2016} and references therein; \citealt{Benitez-Llambay2020}).

A challenge that cannot be accounted for by baryonic effects is the planar distribution of the satellite galaxies around the Milky Way (MW). \citet{Lynden1976} first pointed out that the satellite galaxies of the MW appear to lie in the same polar great circle as the Magellanic Stream. \citet{Kroupa2005} found that the 11 classical satellite galaxies lie in a highly flattened plane that is oriented almost perpendicular to the disk of the MW. Some of the fainter satellites, globular clusters and streams have been
reported to be also associated with this plane of satellites \citep{Metz2009b,Pawlowski2012}. Subsamples of satellites around Andromeda (M31) also show evidence of disc-like features \citep{Metz2007}. With the discovery of additional satellite galaxies of M31 by the Pan-Andromeda
Archaeological Survey (PAndAS; \citealt{McConnachie2009}), \citet{Conn2013} and \citet{Ibata2013} identified a plane that consists of 15 out of the 27 dwarf galaxies around M31. Such planar distributions of satellite galaxies are also found outside the Local Group. \citet{Tully2015} reported two parallel planes of satellites in the Centaurus A Group that was later revised by \citet{Muller2018} to be one single structure. Such highly flattened distributions of satellite galaxies are not typical in $\Lambda$CDM simulations, which generally predict less anisotropic distributions of substructures \citep{Libeskind2005}. Estimates of the probability of finding such highly flattened distributions of satellite galaxies range from a few per cent \citep{Wang2013,Pawlowski2014,Shao2016} to 10 per cent when factors such as the `look elsewhere' effect are taken into account \citep{Cautun2015b}.

The anisotropic distribution of satellite galaxies could be related to the dark matter (DM) halo properties. \cite{Libeskind2005},  \cite{Zentner2005} and \cite{Shao2020} found that the long axis of the  elongated disk composed of subhaloes aligns with the major axis of their host halo. The halo spin and shape and the orbital angular momentum of the subhaloes are correlated with the large-scale structure (LSS) \citep{Colberg2005,Kasun2005,Altay2006,Zhang2009,Paz2011,Libeskind2013a,Shao2019}, and thus the LSS could also be responsible for the anisotropic distribution of satellite galaxies. Indeed, multiple previous works have shown that the anisotropic distribution of satellite systems could be due to the preferential infall of satellites (or subhaloes) along the spine of filaments \citep{Zentner2005,Libeskind2005,2011,Lovell2011,Buck2015,Ahmed2017}.

Alternatively, \citet{Li2008} showed that if all of the 11 MW satellite subhaloes were accreted in a single group, the probability of a disk-like structure in the satellites is enhanced. However, they did not investigate the likelihood of all the 11 brightest satellites being members of one group. \citet{Metz2009a} has shown that group accretion is inconsistent with the observed properties of dwarf galaxy groups. \citet{Wang2013} showed that only 30 per cent of the top 11 satellites in the Aquarius simulations share the same friends-of-friends group before infall. Recent work also showed that the 11 most massive satellites of MW-mass halos are mostly
(75 per cent) accreted individually, 14 per cent in pairs and 6 per cent in triplets, with higher group multiplicities being very unlikely \citep{Shao2018}.

Observational studies of the distribution of satellite galaxies in MW analogues are difficult to perform since their satellite galaxies are usually too faint to be detected, except in the closest such systems. It is more feasible observationally to extend the study to galaxy clusters to explore whether such highly flattened distributions of satellite galaxies exist in these high mass systems. Previous works showed that satellite galaxies exhibit anisotropic distributions and are preferentially located along the major axis of the brightest cluster galaxy \citep[e.g.][]{Carter1980, Yang2006, Wang2008, Niederste-Ostholt2010, Hao2011, Huang2016}. Such anisotropy and alignments are also reported in cosmological simulations \citep[e.g.][]{Kang2007, Ragone-Figueroa2020}. \cite{Paz2006} found a strong dependence of the satellite distribution on cluster mass. The higher the mass, the larger the triaxiality parameter of the satellite distribution is. 

The anisotropic distributions could be related to the anisotropic shapes of DM halos \citep[e.g.][]{Shin2018}. Cosmological simulations show that the halos of galaxy clusters are triaxial rather than spherical \citep[e.g.][]{Frenk1988, Jing2002, Cuartas2011, Bonamigo2015, Vega-Ferrero2017} and their ellipticity increases with cluster mass \citep{Kasun2005, Bailin2005, Hopkins2005, Despali2014, Despali2017, Okabe2020}. The elliptical shapes of DM halos have also been found using gravitational lensing \citep[e.g.][]{Oguri2010,Gonzalez2021}, SZ-effect \citep[e.g.][]{Filippis2005,Sayers2011} and X-ray studies \citep[e.g.][]{Kawahara2010,Sereno2013}.

In this paper, we use semi-analytical galaxy formation models to explore whether thin planes of satellites should exist in galaxy clusters and how the spatial distribution of satellite galaxies is related to the properties, formation history and large-scale environment of a galaxy cluster. We also study satellite galaxy distributions at accretion. The paper is organized as follows. In Section \ref{sec:data and methods}, we introduce the simulations and methods used in this study. We present our results in Section \ref{sec:result} and our conclusions in Section \ref{sec:conclusion}. 

\section{DATA AND METHODS}
\label{sec:data and methods}
\subsection{Simulations}
\label{sec:simulation}

We use the simulated galaxy catalogue of \citet{Guo2013} based on the Millennium-WMAP7 simulation (MS7; \citealt{Springel2005}). The MS7 is a cosmological simulation of a periodic cube of 500 comoving $h^{-1}$Mpc ($h^{-1}$cMpc) side length and follows the evolution of 2160$^{3}$ DM particles with a particle mass of $9.3639\times10^{8}$ $h^{-1} \rm M_{\odot}$ from redshift 127 to the present day. The cosmological parameters of the MS7 are consistent with the the seven-year \emph{Wilkinson Microwave Anisotropy Probe (WMAP)} results: $\Omega_{\rm m}=0.272$, $\Omega_{\rm b}=0.0455$, $\Omega_{\Lambda}=0.728$, $h=0.704$, $\sigma_8=0.81$, and $n=0.967$.

We also use the scaled Millennium-II simulation (MS-II; \citealt{Boylan2009}) to select a comparison sample of MW analogues. The original MS-II adopted the cosmological parameters of the first-year \emph{WMAP} results: $\Omega_{\rm m}=0.25$, $\Omega_{\rm b}=0.045$, $\Omega_{\Lambda}=0.75$, $h=0.73$, $\sigma_8=0.9$, and $n=1$. \cite{Guo2013} has scaled it to the seven-year \emph{WMAP} parameters with the technique described in \citet{Angulo2010}. The rescaled MS-II corresponds to a simulation following $2160^3$ particles in a periodic cube of 104.3 $h^{-1}$cMpc on a side, with each DM particle having a mass of $8.50\times10^{6}$ $h^{-1} \rm M_{\odot}$. Hereafter we refer to the scaled \emph{WMAP7} version of the MS-II as MSIIsc7.

A standard friends-of-friends (FOF; \citealt{Davis1985}) algorithm with a linking length of 0.2 times the mean interparticle separation is used to identify FOF groups. A minimum number of 20 particles is imposed for each FOF group. For each FOF group, the virial radius, $R_{\rm vir}$ is defined as the radius of the sphere centred at the potential minimum of the FOF group within which the average density is 200 times the critical density of the universe. The mass inside the virial radius is defined as the virial mass, $M_{\rm vir}$. Subhaloes are identified in each FOF group using the SUBFIND algorithm \citep{Springel2001}. The main subhalo centred at the potential minimum of the FOF group is assigned the FOF virial mass and virial radius. Galaxies residing in main subhaloes are referred to as central galaxies.

Galaxy catalogues are generated by implementing the \cite{Guo2013} semi-analytical galaxy formation model on the merger trees of the MS7 and MSIIsc7. This semi-analytical model adopts various prescriptions describing the relevant processes of galaxy formation, including gas infall, cooling, star formation, supernova feedback and AGN feedback, galaxy mergers and metal enrichment. It has proven successful in reproducing many galaxy properties both in the Local Universe and at high redshift \citep{Xie2015,Guo2015,Buitrago2017,Rong2017a,Rong2017b}. Readers are referred to the original papers for detailed descriptions of the simulations and galaxy formation models. 

In order to remove any mass dependence, we select galaxy clusters in the MS7 within a narrow mass range: $M_{\rm vir}\in{}\left(1,\ 3\right)\times10^{14}$ $\rm M_{\odot}$, which leads to 2587 galaxy clusters. Given the mass resolution of the simulation and completeness requirement in the observations, only galaxies with stellar masses larger than 10$^{9.5}$ $\rm M_{\odot}$ are used for the analysis. Galaxies within 1 Mpc from the central galaxies are classified as satellite galaxies for the 3D analysis. Each cluster has 38 satellite galaxies on average, with a standard deviation of 10 and has at least 11 satellites. We study the distributions of the subsystems of the 11 most massive satellites to have a fair comparison with the MW system  and to have complete samples with enough statistics both in terms of satellite numbers and system numbers. The identification of satellite galaxies in the 2D analysis is identical to that described in Section \ref{sec:obsdata} for observational data.

To identify MW-mass halos we adopt the same selection criteria as \citet{Shao2019}. We select halos with virial mass
$M_{\rm vir}\in{}\left(0.3,\ 3\right)\times10^{12}$ $\rm M_{\odot}$ from the MSIIsc7 and further require the halos to be isolated by removing those with central galaxies that have a neighbour more massive in stars than half of the central galaxy within 600 kpc. We also require that each central galaxy be accompanied by at least 11 satellite galaxies within a distance of 300 kpc from the central galaxy with stellar masses larger than $10^6$ $\rm M_{\odot}$. The resolution of the MSIIsc7 is high enough to analyze these satellite galaxies around MW analogues. \cite{Guo2011} shows that it can reproduce the abundance of the MW satellite galaxies as a function of V-band magnitude up to -5. \cite{Wang2013} proves the 11 most massive satellite galaxies in MW analogues predicted by the MS-II follow the same radial profile as observed. \cite{Cautun2015b} shows that the spatial distributions of these satellite galaxies are consistent with those predicted in the \emph{Copernicus Complexio} (COCO; \citealt{Hellwing2016}) simulation, a simulation with 60 times higher mass resolution. We thus trust the galaxies down to $10^6 \rm M_{\odot}$. Finally, we have 4405 MW analogues.

\subsection{Observational data}
\label{sec:obsdata}

 We use the galaxy group catalogue constructed by \citet{Yang2007} which is  based on the Seventh Data Release of the Sloan Digital Sky Survey (SDSS DR7; \citealt{Abazajian2009}) and the New York University Value-Added Galaxy Catalog (NYU-VAGC; see \citealt{Blanton2005}). They selected all galaxies in the main galaxy sample with redshifts between 0.01 $\leq$ $z$ $\leq$ 0.20, with redshift completeness, $\mathit{C}$ > 0.7,  and SDSS $r$-band magnitude, $r_{\rm mag} < 17.77$. Similar to our simulated sample, we select clusters which have virial mass $M_{\rm vir}\in{}\left(1,\ 3\right)\times10^{14}$ $\rm M_{\odot}$. We adopt the stellar mass from the MPA-JHU (Max Planck Institute for Astrophysics and the Johns Hopkins University) DR7 catalogue  (\citealt{Kauffmann2003}; \citealt{Brinchmann2004}). Central galaxies are defined as the most massive galaxies in the corresponding clusters in the group catalogue. We define satellite galaxies by requiring that: (i) the projected distance to the central galaxy is 0.1 Mpc < $r_{\rm p}$ < 1 Mpc, and (ii) the line-of-sight velocity difference from the central galaxy is |$\Delta v$| < 1000 km s$^{-1}$. We remove satellite galaxies within 0.1 Mpc to avoid fibre collision effects. In total, we have 516 galaxy clusters with at least 11 satellite galaxies in the SDSS.

\subsection{Projected distribution}
\label{sec:ellipticity}
To make a direct comparison with observations, we calculate the projected 2D ellipticity of the 11 presently most massive satellites system for each cluster using the method of \citet{Evans2009}. The quadrupole moments are given by 

\begin{subequations}
\begin{align}
    Q_{\rm xx} & =\langle(x_{i}-x_{c})^2\rangle_{i} \ , \\
    Q_{\rm xy} & =\langle(x_{i}-x_{c})(y_{i}-y_{c})\rangle_{i} \ , \\
    Q_{\rm yy} & =\langle(y_{i}-y_{c})^2\rangle_{i} \ ,
\end{align}

\end{subequations}
where the summation $i$ is over the 11 most massive satellites. The ellipticity components, $e_1$ and $e_2$, of the satellite distribution are obtained through the equations

\begin{subequations}

 \begin{align}
    e_{\rm 1} & =\frac{Q_{\rm xx}-Q_{\rm yy}}{Q_{\rm xx}+Q_{\rm yy}+2\sqrt{Q_{\rm xx}Q_{\rm yy}-Q_{\rm xy}^2} } \ , \\
    e_{\rm 2} & =\frac{2Q_{\rm xy}}{Q_{\rm xx}+Q_{\rm yy}+2\sqrt{Q_{\rm xx}Q_{\rm yy}-Q_{\rm xy}^2}} \ .
 \end{align}
 
\end{subequations}

The overall ellipticity is defined as
\begin{equation}
e=\sqrt{e_{\rm 1}^2+e_{\rm 2}^2}
\end{equation}
The ellipticities vary from zero (circular) to unity (linear).

\subsection{3D distributions}
\label{sec:3d distribution}
One of the advantages of the simulated galaxy catalogue is that it provides the 3D distribution of galaxies. In the following sections, we define the 3D shape of the satellite distribution, the host halo and the LSS and quantify the degree of alignment between these systems.

\subsubsection{Eigenvectors and eigenvalues}
\label{sec:axis ratio}
We first calculate the mass tensor
\begin{equation}
I_{ij} \equiv \sum_{k=1}^{N} x_{k,i} x_{k,j} \ ,
\end{equation}
where $N$ is the number of members in a given system and $x_{k,i}$ denotes the $i$th position component ($i =$ 1, 2, 3) of the $k$th member with respect to its centre. For the satellite system, the sum is over its satellite galaxies. For the host halo, the sum is over all DM particles within $R_{\rm vir}$, where we fix $R_{\rm vir}=1$ Mpc for clusters. For the LSS, the sum is over all DM particles within the spherical shell located between $2R_{\rm vir}$ and $3R_{\rm vir}$ from the centre of the host halo, where the $R_{\rm vir}$ is the virial radius of the host halo.

The shape and orientation are determined by the three eigenvalues, $\uplambda_{i}$ ($i$ $=$ 1, 2, 3, $\uplambda_1\geq\uplambda_2\geq\uplambda_3$), and the normalized eigenvectors, $\boldsymbol{\hat{e}}_{i}$, of the mass tensor. The major, intermediate and minor axes of the ellipsoid are given by $a$ $=$ $\sqrt{\uplambda_1}$, $b$ $=$ $\sqrt{\uplambda_2}$ and $c$ $=$ $\sqrt{\uplambda_3}$, respectively. The orientation is defined as the direction of the minor axis, $\boldsymbol{\hat{e}}_3$, for the relevant system.

\subsubsection{Thickness of the plane of satellites}
\label{sec:thickness}

We adopt an alternative way to describe the flatness of the satellite galaxy distributions \citep{Kroupa2005}, calculating the best-fitting plane by minimizing the root-mean-square ($\rm rms$) of the height of each satellite related to the plane of satellites. The thickness of the plane is defined as the ratio of the $\rm rms$ height to the virial radius of the DM halo,  $\tilde{h}_{\rm thick} \equiv h_{\rm rms}/ R_{\rm vir}$, where we fix $R_{\rm vir}$ $=$ 1 Mpc for clusters and $R_{\rm vir}$ $=$ 0.3 Mpc for MW analogues. The height, $h_{\rm rms}$, is thus given by 

\begin{equation}
    h_{\rm rms} \equiv \sqrt{\frac{\sum_{i=1}^N (\boldsymbol{n}\cdot \boldsymbol{x}_{i})^2}{N}} \ .
\end{equation}
Here $N$ is the number of satellites and $\boldsymbol{n}$ is the normal direction of the plane. $\boldsymbol{x}_{i}$ denotes the position (in units of Mpc) of each satellite with respect to the host centre.

\subsubsection{Alignment}
\label{sec:angle}
We will study the degree of alignment between the satellite distributions, host halos and LSS, denoted by the misalignment angle, $\rm \theta$. For example, the misalignment angle between the plane of satellites and the host halo is defined as,
\begin{equation}
  \rm  \theta_{Sat-Halo}=arccos( |\boldsymbol {\hat n}^{\rm Sat} \cdot \boldsymbol {\hat e}_3^{\rm Halo}| ) \ ,
\end{equation}
where $\boldsymbol {\hat n}^{\rm Sat}$ and $\boldsymbol {\hat e}_3^{\rm Halo}$ are the orientations of the plane of satellites and the host halo, respectively. The misalignment angles between the plane of satellites and the specific angular momentum of the host halo, $\rm \theta_{Sat-Spin}$, between the plane of satellites and the LSS, $\rm \theta_{Sat-LSS}$, and between the host halo and the LSS, $\rm \theta_{Halo-LSS}$, are calculated in the same way.

\subsection{Prominence}
\label{sec:prominence}

 Quantifying the degree of anisotropy as the ellipticity, $e$, or the flattening, $c/a$, and the fractional thickness, $\tilde{h}_{\rm thick}$, of the satellite system introduces an unwanted dependence on the radial distribution of satellites, as we shall see in the next section. This dependence makes it challenging to compare the degree of anisotropy between clusters and MW-mass hosts since these populations have somewhat different radial distributions of satellites. One approach to mitigate this effect has been proposed by \cite{Cautun2015b}; (see also \citealt{Libeskind2005}) and consists of quantifying the degree of anisotropy as the probability of obtaining a given satellite system from random fluctuations of an isotropic distribution with the same radial distribution as the sample of interest. For each satellite system, we generate $10^4$ isotropic satellite samples by fixing the radius of each satellite galaxy and randomizing its position angle with respect to the centre. Here we do not use the information of the host halo shape, which is unusually triaxial. We test such an effect by multiplying the halo axis ratios for the random samples and find it does not change our results qualitatively.

The prominence, $\mathcal{P}_{e}$, is then defined as
\begin{equation}
    \mathcal{P}_{e} \equiv \frac{1}{ p(\geq{[e]_{i}})}
\end{equation}
where $[e]_{i}$ is the measured ellipticity value and $p(\geq{[e]_{i}})$ is the probability of an isotropic system to have $e \geq [e]_{i}$. The larger the prominence, the less likely is it that system of satellites originates from a random distribution. In a similar way, we define the prominence of a given axis ratio and thickness as, $\mathcal{P}_{c/a} \equiv 1/ p(\leq{[c/a]_{i}})$ and $\mathcal{P}_{\tilde{h}_{\rm thick}} \equiv 1/p(\leq{[\tilde{h}_{\rm thick}]_{i}})$.

\section{Results}
\label{sec:result}

In this section, we first compare the 2D ellipticity of satellite galaxy distributions between the simulations and the SDSS. Then we use the 3D satellite distributions in the MS7 to investigate the correlation of the spatial distributions of satellite galaxies with their host halos, as well as with the LSS environment. We further trace the galaxy merger trees to study the effect of accretion and the contribution of group accretion. Throughout this section, we refer to the 11 most massive satellite galaxies as the top 11 satellite galaxies.

\subsection{Projected distribution of satellite galaxies around clusters}
\label{sec:3.1}
  
\begin{figure}
\centering
\includegraphics[width=0.98\columnwidth]{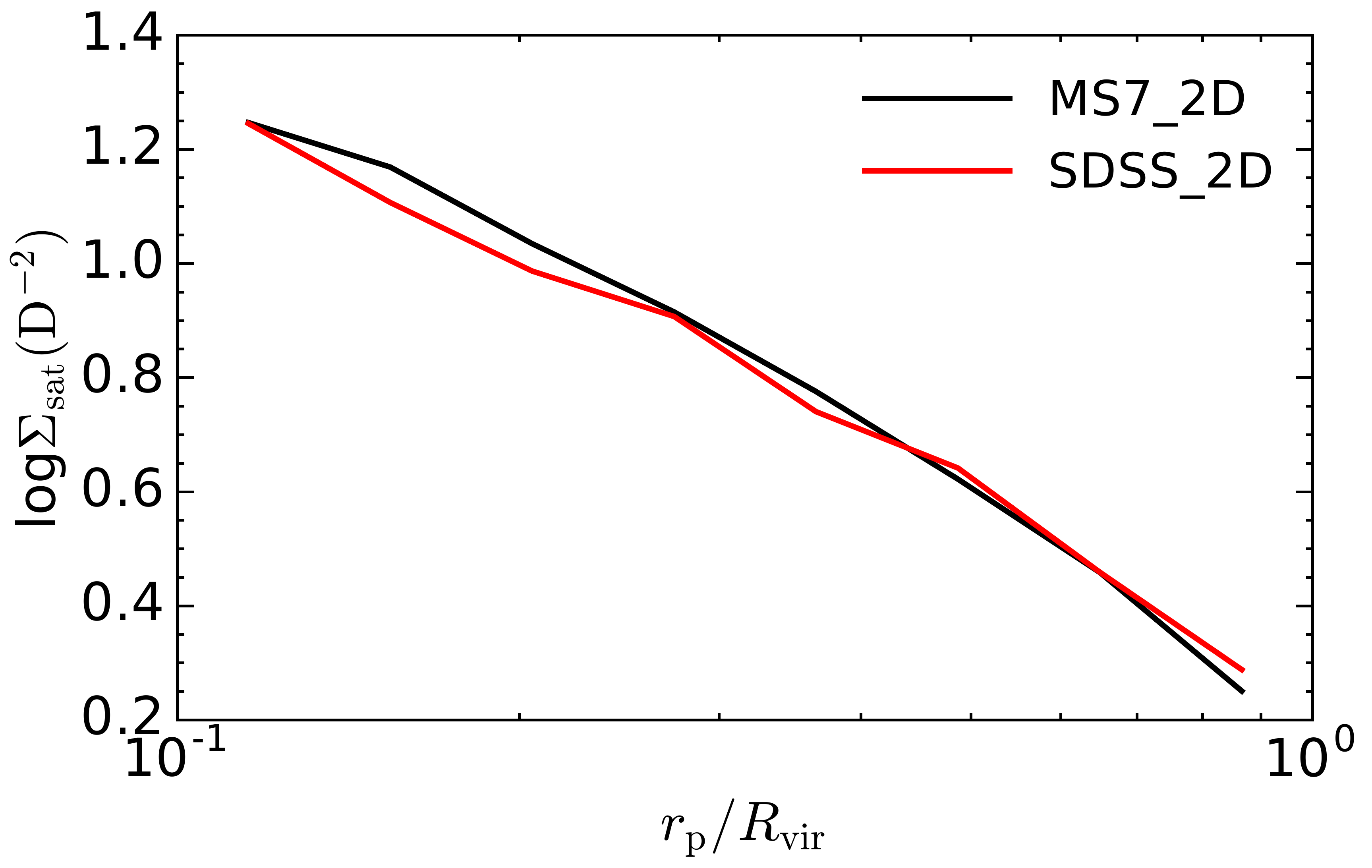}
\includegraphics[width=0.98\columnwidth]{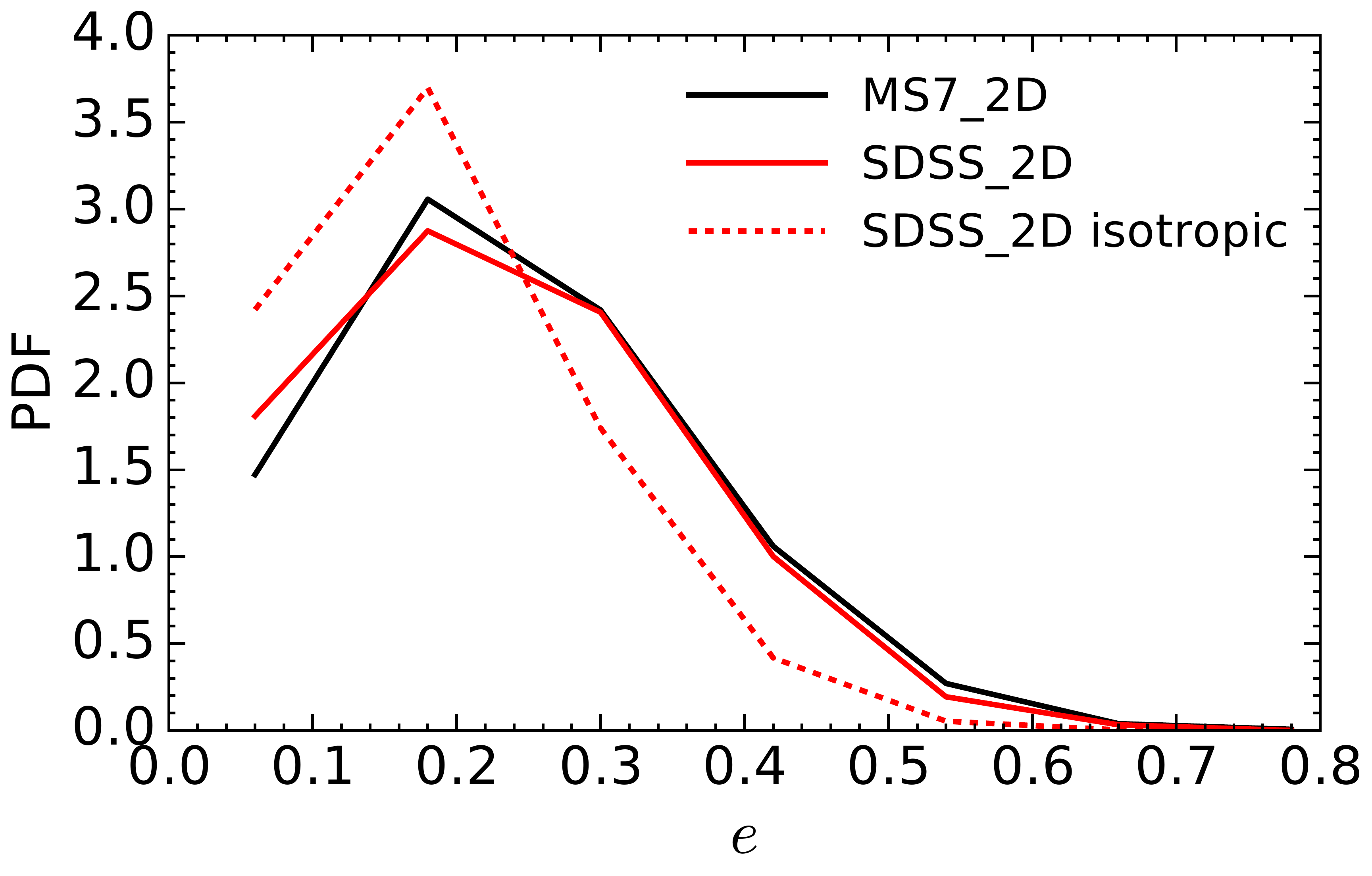}
\includegraphics[width=\columnwidth]{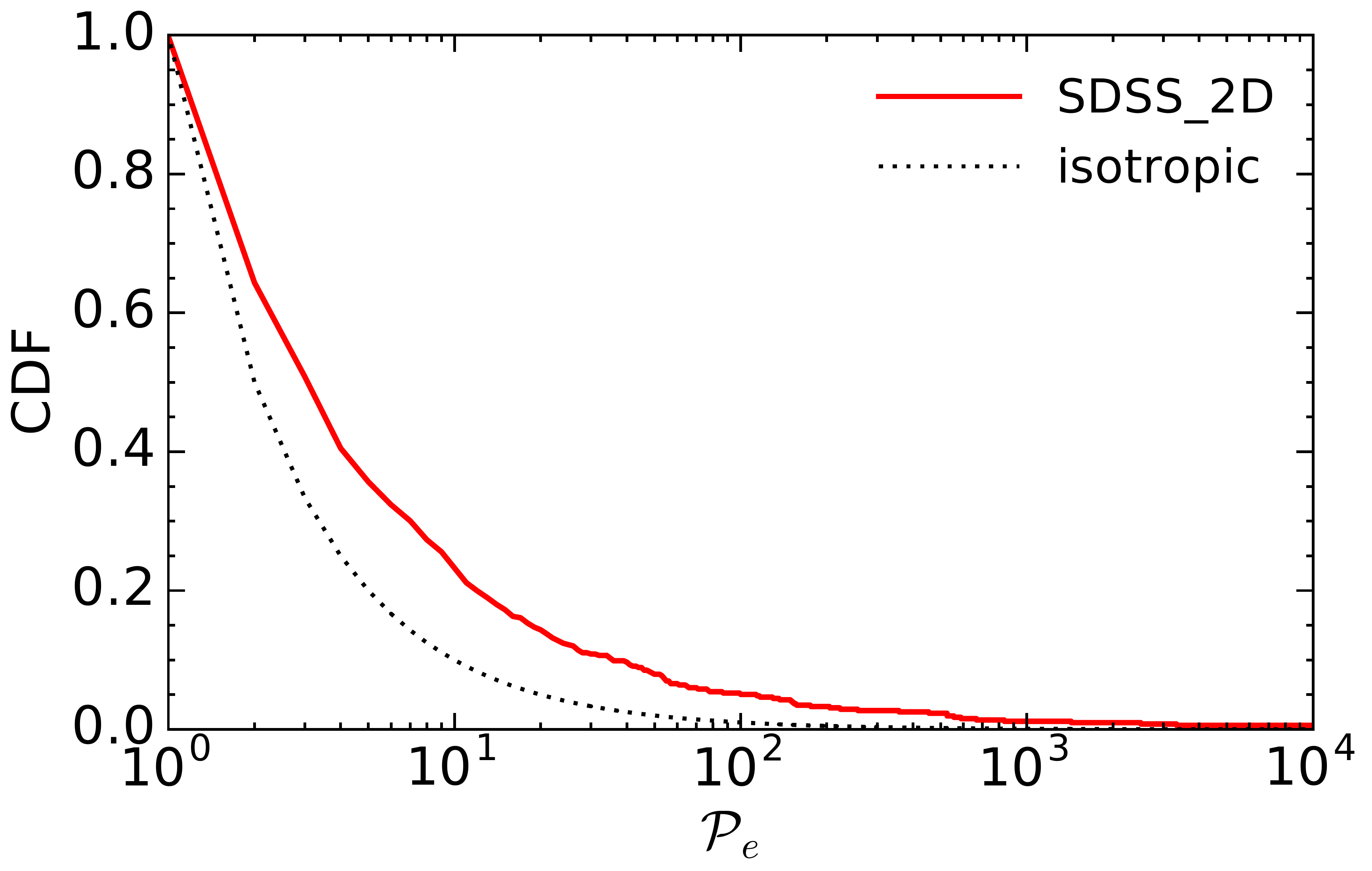}
\caption{Comparison of the subsystems of the 11 satellites with highest stellar mass of galaxy clusters in the SDSS and the MS7 mock catalogue. \textit{Top panel}: the average projected number density profiles. The profiles give the satellite number counts per unit surface area in units of $D \equiv r_{\rm p}/R_{\rm vir}$. \textit{Middle panel}: the probability density function (PDF) of the ellipticity of these subsystems. The black solid curve shows the result in the MS7 mock catalogue, while the red solid curve shows the observed result using the catalogue of \citet{Yang2007}. The result based on the isotropic sample is shown as the red dashed line in the SDSS which is nearly the same as the distribution in the MS7 (not shown here). \textit{Bottom panel}: the complementary cumulative distribution function (CDF) of the prominence, $\mathcal{P}_e$, of the ellipticity in the SDSS and isotropic distribution.}
\label{fig_obervation}
\end{figure}

  Among the top 11 satellites in the simulated clusters, there are about 50\% orphan galaxies,  which lost their dark halos either due to the stripping processes in the dense environment or due to the numerical effects. The positions of orphan galaxies are then calculated assuming dynamical friction-induced evolution. In addition, satellite galaxies are hosted in subhalos of relatively shallower potentials and are thus more sensitive to baryonic processes. 
  Some members of the 11 most massive satellite galaxies could switch with other satellite galaxies once their stellar masses change if invoking different galaxy formation models and parameters. In order to test whether such treatments are reliable, we compare the satellite distribution with observations. For it is difficult to obtain the 3D distributions of observed satellite galaxies, in this section, we focus on the projected distributions of satellite galaxies.

  To compare with the SDSS, we generate a mock catalogue by assigning each galaxy a redshift based on its line-of-sight distance and peculiar velocity assuming an observer at the origin of the coordinates. In practice, we define the redshift of a given galaxy as $z_{gal} = H_0\it d_{\rm gal}/c+\it v_{\rm p}/c$, where $H_0 = 70.4\ \rm kms^{-1}Mpc^{-1}$, $d_{\rm gal} = (x^2+y^2+z^2)^{1/2},\ c = 3 \times10^8$ m/s and $v_{\rm p}$ is the peculiar velocity along the line-of-sight direction. For each cluster satisfying $M_{\rm vir}\in{}\left(1,\ 3\right)\times10^{14}$ $\rm M_{\odot}$, the central galaxy is the one associated with the main subhalo and we identify its satellite galaxies according to their projected radii and redshifts as described in Section \ref{sec:obsdata}. Such selection criteria are similar but not identical to those in Section \ref{sec:obsdata} that we do not adopt the identical central galaxy identification and the luminosity restriction. We have tested the effect of such differences in an available light-cone mock catalogue based on the Millennium simulation with somehow different cosmological parameters and have found that these differences do not change the results significantly. 
  
   The projected number density profiles of the top 11 satellites in the mock catalogue of the MS7 and that in the SDSS are shown in the top panel of Fig. \ref{fig_obervation}. This clearly demonstrates that the radial distribution of satellites in our mock galaxy catalogue is consistent with that observed (see also \citealt{Guo2011}). The distributions of the ellipticity of the top 11 satellites system agree well between the mock catalogue and the SDSS as well, as shown in the middle panel of Fig. \ref{fig_obervation}. A two-sample Kolmogorov-Smirnov (KS) test on the ellipticity distributions between the SDSS and the mock catalogue of the MS7 gives a $p$-value of 0.28, which indicates that any differences are consistent with statistical noise. The good agreement between the mock catalogue and the SDSS allows us to study the satellite distribution using this simulation in more detail.
  
  The ellipticity in the SDSS has a mean value of 0.23 $\pm$ 0.12, similar to the result of 0.21 $\pm$ 0.11 in \cite{Huang2016} and 0.21 $\pm$ 0.04 in \cite{Gonzalez2021}. Both in the SDSS and in the mock catalogue of the MS7 the top 11 satellite systems have, on average, higher ellipticity, i.e. they are more anisotropic than expected due to random fluctuations. We quantify the excess of anisotropy using the prominence distribution shown by the red solid curve in the bottom panel of Fig. \ref{fig_obervation} (see Section \ref{sec:prominence}). This is consistent with the results found in previous studies \citep{Clampitt2016,Shin2018}.

\begin{table}
    \centering
    \caption{The fraction of simulated systems whose flattening (as measured by $c/a$ or $\tilde{h}_{\rm thick}$) is less than the [16, 50, 84] percentiles for the flattening distribution obtained in the isotropic case. The first column gives the name of the properties and systems, the next three columns give the fractions of systems that are more flattened than the corresponding percentiles of the isotropic distribution.
    All properties in simulations are measured at $z=0$ except for the last two rows which are measured at accretion and at $z=0.8$, respectively. The isotropic samples have the same radial distribution as the simulated satellites and are obtained by randomizing the position angles.} 
    \label{tab:fraction}
    \small
    \setlength{\arrayrulewidth}{0.2mm}
    \setlength{\tabcolsep}{2.9pt}
    \renewcommand{\arraystretch}{1.5}
   \begin{tabular}{lccc}
        \hline
        Property & 16th percentile & 50th percentile & 84th percentile \\
        \hline \hline

        $c/a$ clusters & 42.8\% & 74.4\% & 93.0\% \\
        $c/a$ MW-mass & 31.7\% & 65.4\% & 90.1\% \\
        $\tilde{h}_{\rm thick}$ clusters & 44.6\% & 76.0\% & 94.6\% \\
        $\tilde{h}_{\rm thick}$ MW-mass& 38.7\% & 72.3\% & 92.6\% \\
       $c/a$ clusters at $z=z_{\rm a}$& 84.2\% & 95.6\% & 99.0\% \\
       $c/a$ clusters at $z=0.8$& 58.0\% & 85.0\% & 97.0\% \\
        \hline
    \end{tabular}
\end{table}

\subsection{The 3D spatial distribution of satellite galaxies in clusters }
\label{sec:3d shape}

\begin{figure}
\centering
\includegraphics[width=\columnwidth]{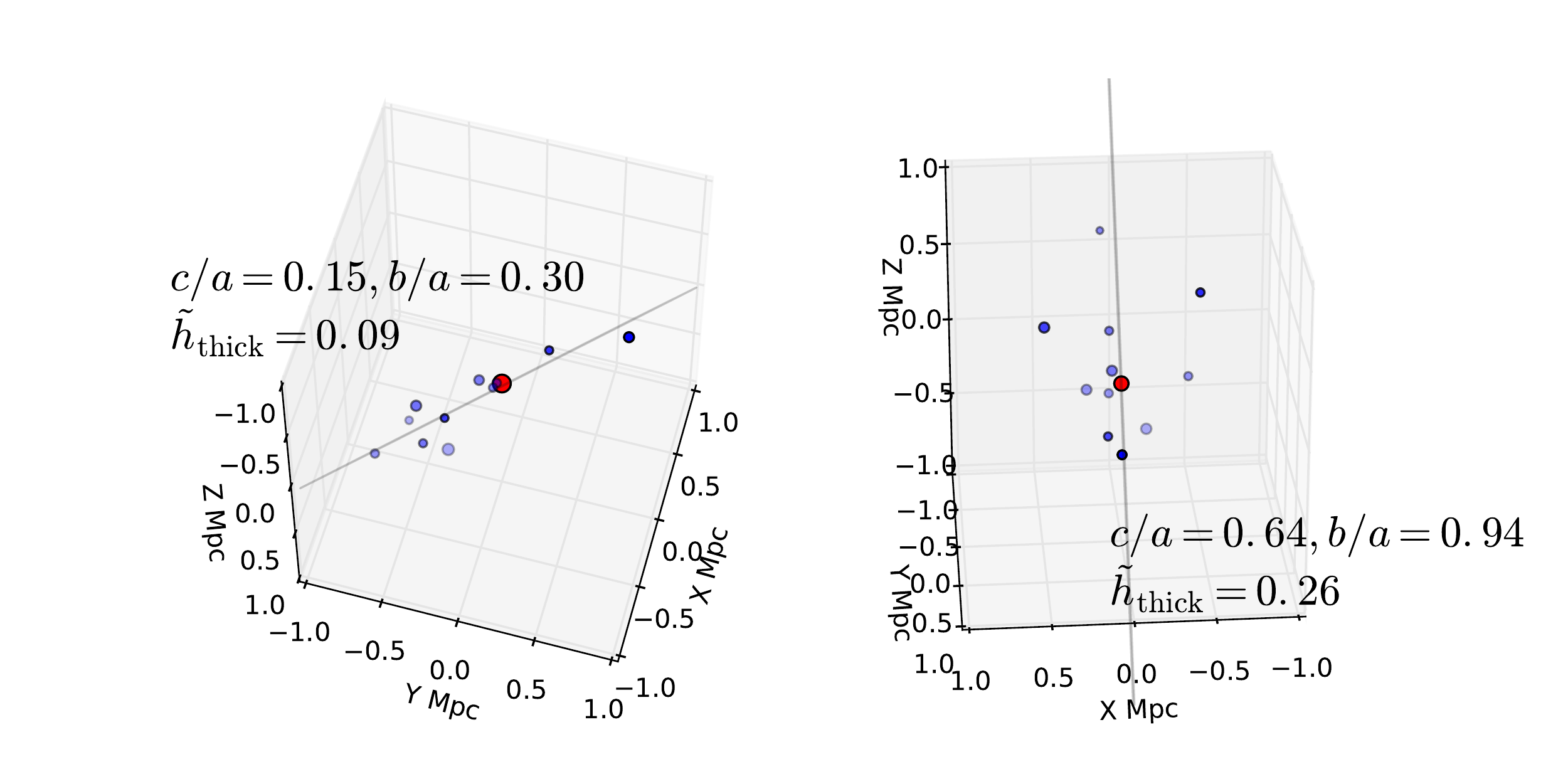}
\includegraphics[width=\columnwidth]{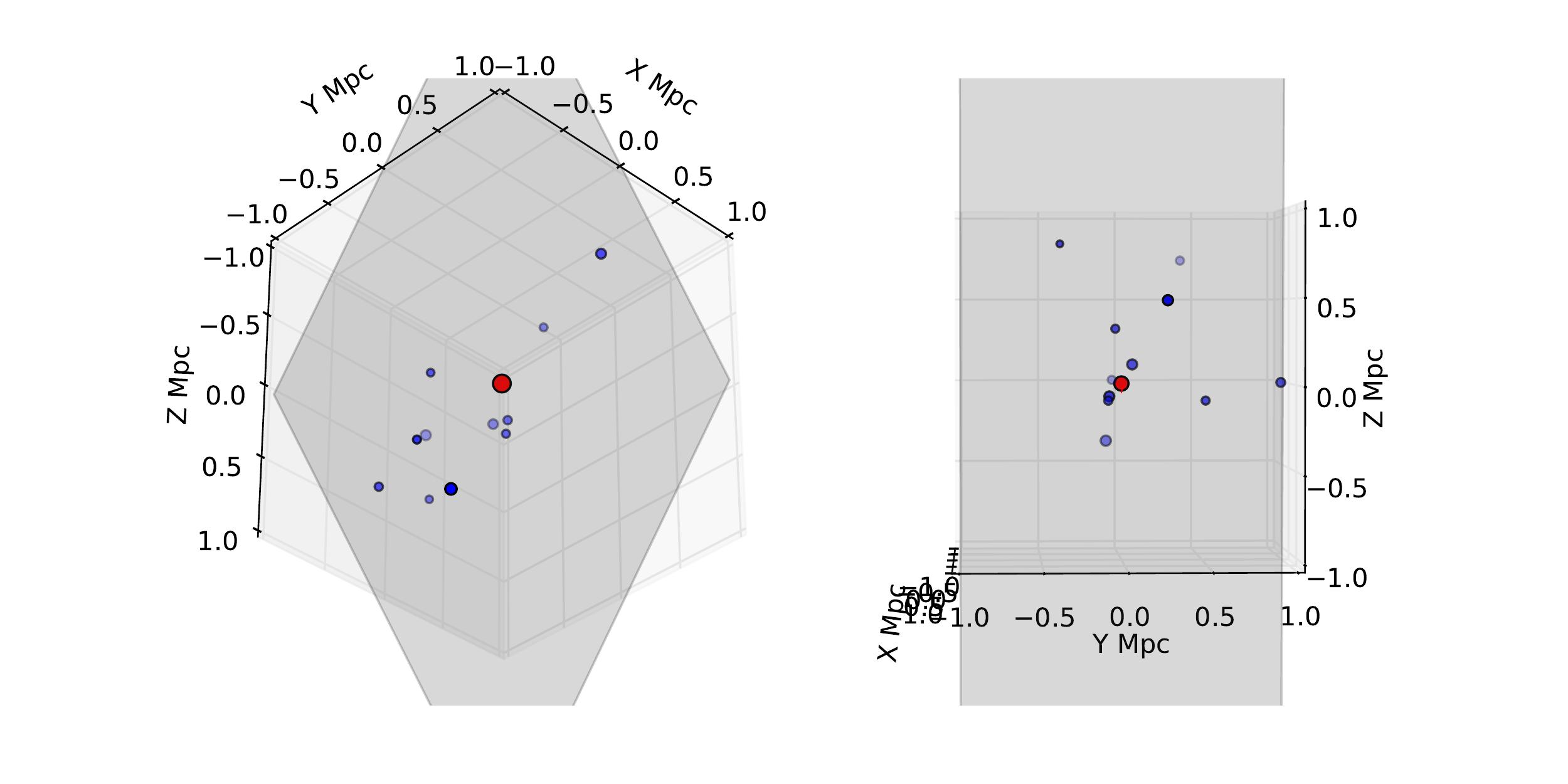}
\caption{Two examples of cluster-mass hosts and the spatial distribution of their 11 satellites with the highest stellar mass. \textit{Left panels}: a simulated galaxy cluster with a thin plane of satellites. \textit{Right panels}: a simulated galaxy cluster with a thick plane of satellites. The top panels are the edge-on view and the bottom panels are the corresponding face-on view. Grey lines and surfaces show the best-fitting planes. Circles are galaxies with sizes proportional to their stellar masses. Red circles represent the central galaxies at $(x, y, z)=0$ and blue circles are the satellite galaxies. The corresponding values of $c/a$, $b/a$ and $\tilde{h}_{\rm thick}$ are given on the top panels.}
\label{shape}
\end{figure}

Although the 3D spatial distribution of satellite galaxies in clusters can reveal more about the formation of the whole system, the measurement requires accurate distances which are not generally available. In the simulations, it is straightforward to study the 3D distribution. Fig. \ref{shape} shows two example satellite distributions, one highly flattened, i.e. small $c/a$, and the other with a nearly isotropic distribution.

\subsubsection{The axis ratio}
\label{sec:c/a}

\begin{figure}
\centering
\includegraphics[width=0.98\columnwidth]{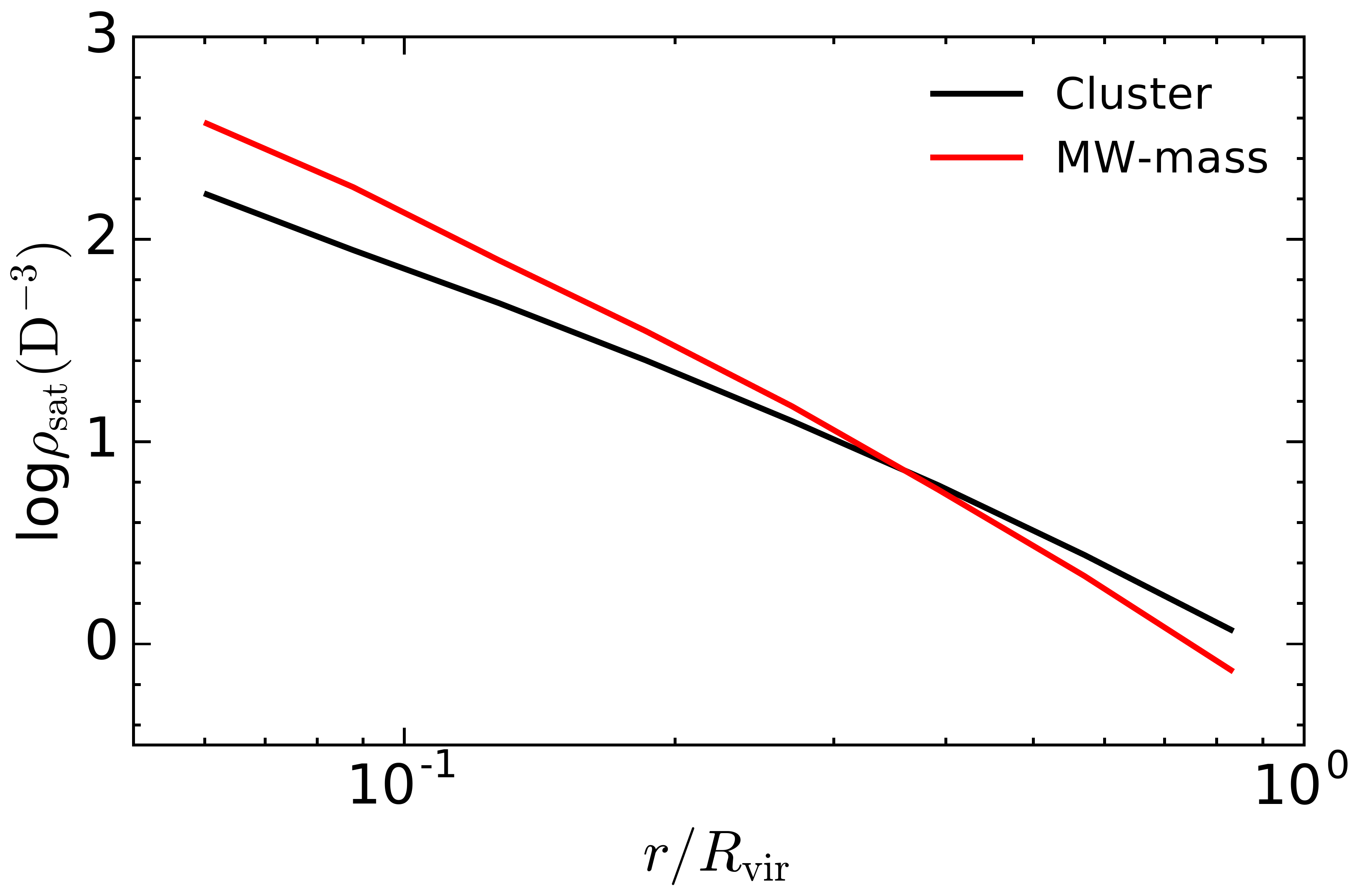}
\includegraphics[width=0.98\columnwidth]{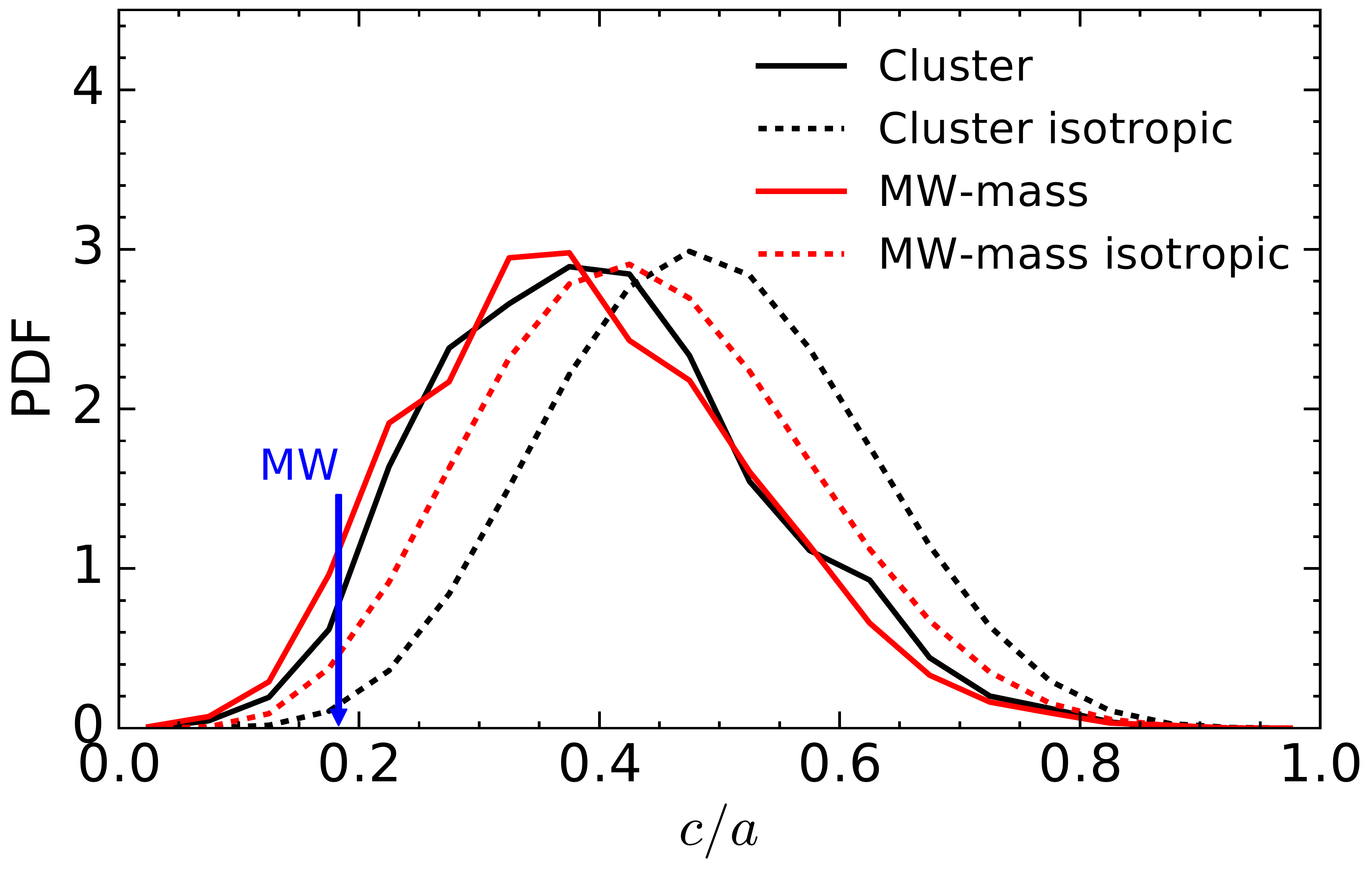}
\includegraphics[width=\columnwidth]{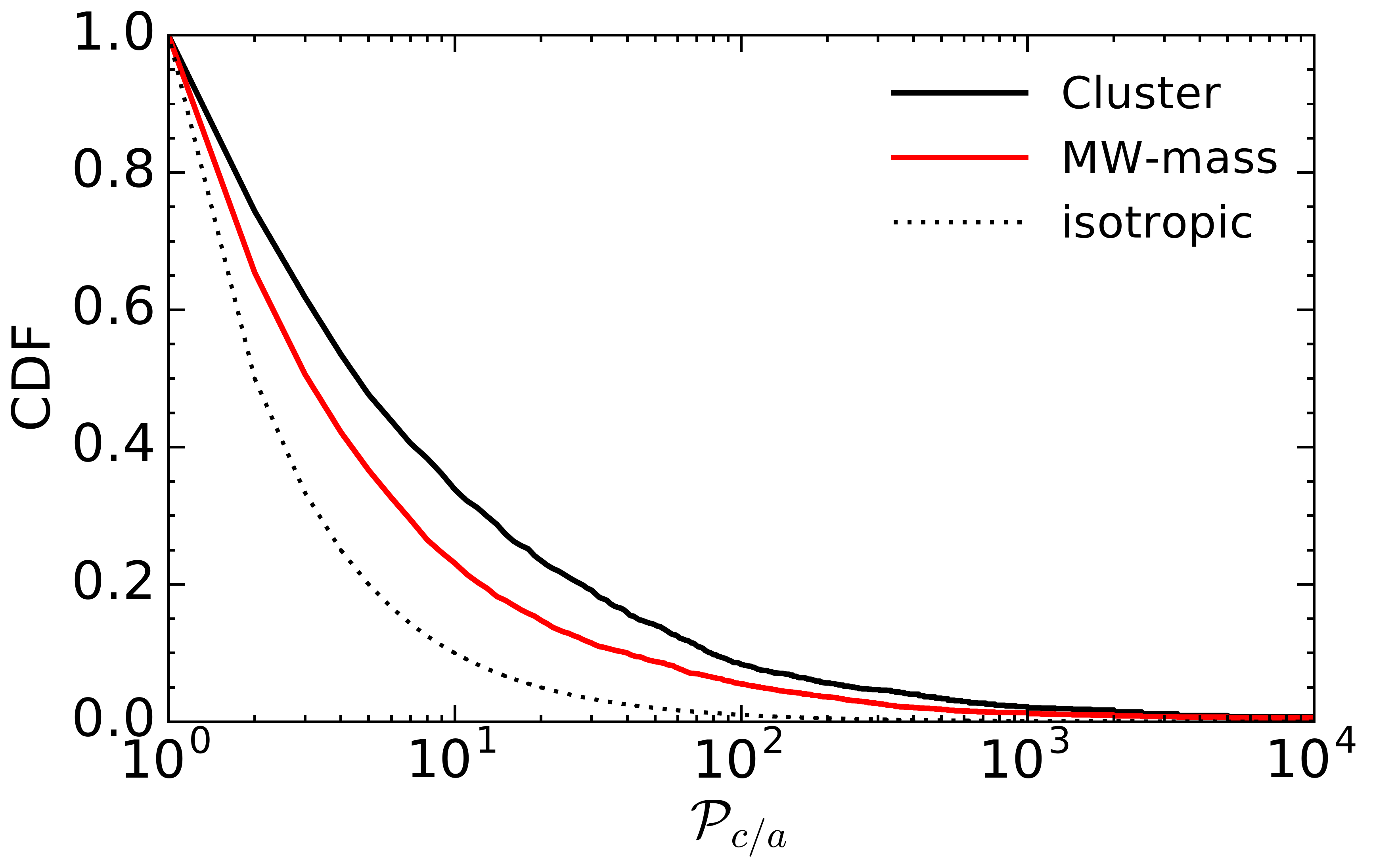}
\caption{\textit{Top panel}: the average 3D number density profiles of the subsystems of the 11 satellites with highest stellar mass in simulated clusters and MW analogues. The profiles correspond to the satellite number counts per unit volume in units of $D \equiv r/ R_{\rm vir}$. \textit{Middle panel}: the minor-to-major axis ratio, $c/a$, of satellite systems in cluster-size halos (black solid) and MW-size halos (red solid). The expectations for isotropic distributions of satellites with the same radial number density as cluster and MW-mass samples are shown with black dashed and red dashed lines. The vertical blue arrow indicates the MW's value, $c/a =$ 0.183. \textit{Bottom panel}: the complementary CDF of the prominence, $\mathcal{P}_{c/a}$, of the axis ratio of satellite systems for clusters (black solid curve) and MW-mass halos (red solid curve). The black dotted curve shows the isotropic distribution.}
\label{c/a_11}
\end{figure}

We study the minor-to-major axis ratio, $c/a$, of the configurations of the top 11 satellite galaxies in our simulated clusters in Fig. \ref{c/a_11}. The top panel shows the number density profiles; the cluster satellites have a less concentrated radial profile (black solid line) than MW analogues (red solid line). We fit the radial distributions with two NFW profiles \citep{Navarro1997} and find that the MW-mass sample has a larger concentration (10.2) than the cluster-mass sample (3.3), as expected from cosmological simulations \citep{Neto2007,Schaller2015,Bose2019}. In the middle panel, the black solid line shows that the $c/a$ of clusters peaks at $\sim$ 0.37, similar to those found in previous results \citep[e.g.][]{West1989,Wang2008}. For each cluster, we generate $10^4$ isotropic samples by fixing the radial distances of the top 11 satellite galaxies but randomizing their angular positions; the result is shown as a black dashed curve in the plot. The isotropic systems have higher $c/a$ ratios than the simulated clusters indicating that the MS7 systems are more flattened than expected due to statistical noise. 

For comparison, we also include the $c/a$ distribution of satellite galaxies in MW analogues. It shows that the cluster and MW-mass satellite systems have similar $c/a$ distributions. For example, 3 per cent (79/2587) of clusters and 5 per cent (211/4405) of MW analogues are flatter than $c/a =$ 0.183 (vertical blue arrow) which is the value of the MW 11 classical satellite galaxies \citep{Shao2016}. \cite{Pawlowski2013} showed the axis ratio of 14 dwarf satellite galaxies of M31 is $c/a=0.125\ \pm\ 0.014$, which is somehow smaller than that of the MW. The close match between the two raw $c/a$ distributions of clusters and MW analogues hides important differences. The cluster satellites are less concentrated than those in MW analogues as shown in the top panel of Fig. \ref{c/a_11}. This lower concentration would lead to a relatively larger $c/a$ for an isotropic distribution in clusters than that in MW-mass systems (shown as black dashed curve and red dashed curve in the middle panel).
On the other hand, previous studies found that the halo shape depends on the halo mass, with massive halos being more anisotropic (smaller $c/a$) \citep{Jing2002,Bailin2005, Allgood2006,Bett2007, Cuartas2011,Despali2017}. One thus expects a smaller $c/a$ in a cluster if satellite galaxies trace the mass distribution. However, satellite galaxies do not necessarily trace the dark matter distribution, and, as we will discuss in detail in Section \ref{sec:relate to halo}, the distribution of satellites tends to be more anisotropic than the host halo. This is consistent with results found for MW-mass analogues \citep{Deason2011}. Taking into account all these factors, it suggests that the close match between the raw $c/a$ distributions of satellite galaxies in clusters and MW analogues just happens by coincidence.

In the bottom panel, we quantify the prominence of the axis ratios of satellite systems in clusters (black solid line) and in MW analogues (red solid line). The prominence of planes of the top 11 satellites is one approach for comparing the degree of anisotropy of systems or populations of systems that have different radial distributions of satellites \citep{Cautun2015b}. This plot shows that the configuration of satellite galaxies deviates more from isotropy in clusters than in MW analogues, which indicates that cluster satellites have a higher degree of anisotropy than MW-mass satellites. The same conclusion can be reached using another test: comparing the fraction of simulated systems that have $c/a$ lower than a certain percentile of their isotropic distribution. For example, Table \ref{tab:fraction} shows that 43\% of clusters and 32\% of MW analogues have $c/a$ lower than the 16 percentiles of their corresponding isotropic distribution. 

\begin{figure}
\centering
\includegraphics[width=\columnwidth]{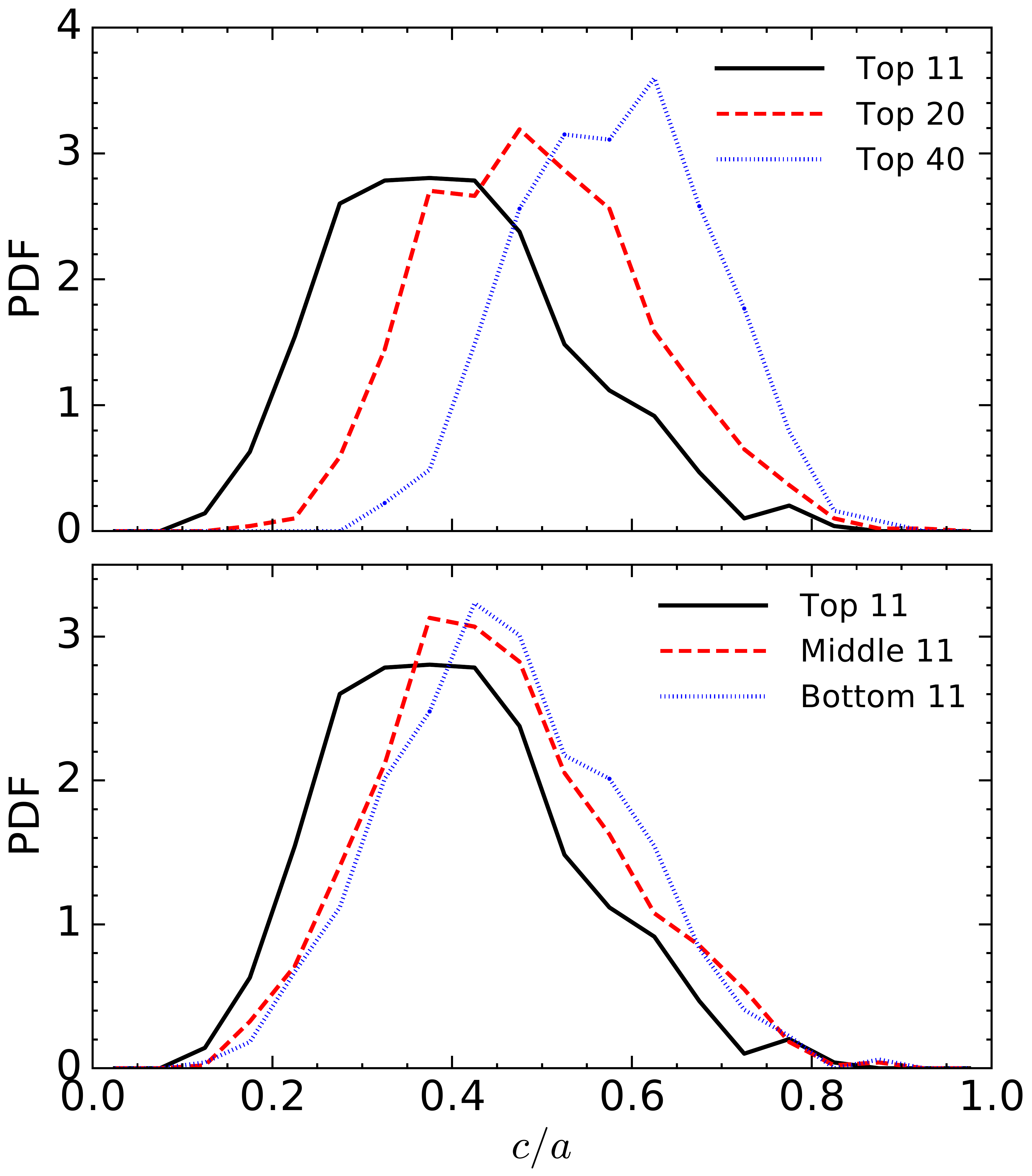}
\caption{The PDF of the minor-to-major axis ratio, $c/a$, of the cluster satellite distributions. We select 984 halos that have at least 40 satellite galaxies each in the MS7. \textit{Top panel}: the $c/a$ distributions of the 11 (black solid), 20 (red dashed) and 40 (blue dotted) most massive satellites, respectively. \textit{Bottom panel}: the $c/a$ distributions of the top (black solid), middle (red dashed) and bottom (blue dotted) 11 satellites galaxies, respectively.}
\label{c/a}
\end{figure}

In order to investigate how $c/a$ varies with the increasing abundance of satellite galaxies, we select 984 cluster-size halos, each containing at least 40 satellite galaxies. We calculate the minor-to-major axis ratios of the systems consisting of the $N=$ 11, 20 and 40 most massive satellite galaxies, respectively. The results are shown in the top panel of Fig. \ref{c/a}. The more satellites are included, the more isotropic the distribution of satellite systems becomes, which is in agreement with the result for MW analogues \citep{Wang2013}. This could be either due to a more isotropic distribution of fainter galaxies, or a reduction of random sampling effects by increasing the number of galaxies, e.g. the axis ratio increases with the increase of the sample size \citep{Pawlowski2017}. We thus check the axis ratios of the top, middle, and bottom 11 massive satellites respectively and find that satellite galaxies with lower masses show a relatively higher $c/a$ as shown in the bottom panel of Fig. \ref{c/a}. The $p$-values of the KS tests on the $c/a$ distributions of the top 11 and middle 11 and of the middle 11 and bottom 11 are $5.39\ \times\ 10^{-12}$ and 0.01, respectively. This suggests a rather different distribution of the subsystem of the 11 satellites with highest stellar mass in comparison with their lower mass counterparts. This is in line with the explanation by \citet{Libeskind2005} that the massive satellites largely preserve the directions in which they were accreted while smaller satellites are often accreted over a longer period of time and from a larger range of directions. The misalignment between the high and low mass satellites further broadens the angular distribution of satellite galaxies, so that when including both massive and less massive satellite galaxies the distribution of $c/a$ tends to be even broader.

\subsubsection{Thickness of the plane of satellites}
\label{sec:rthick/rvir}

\begin{figure}
\begin{center}
\includegraphics[width=\columnwidth]{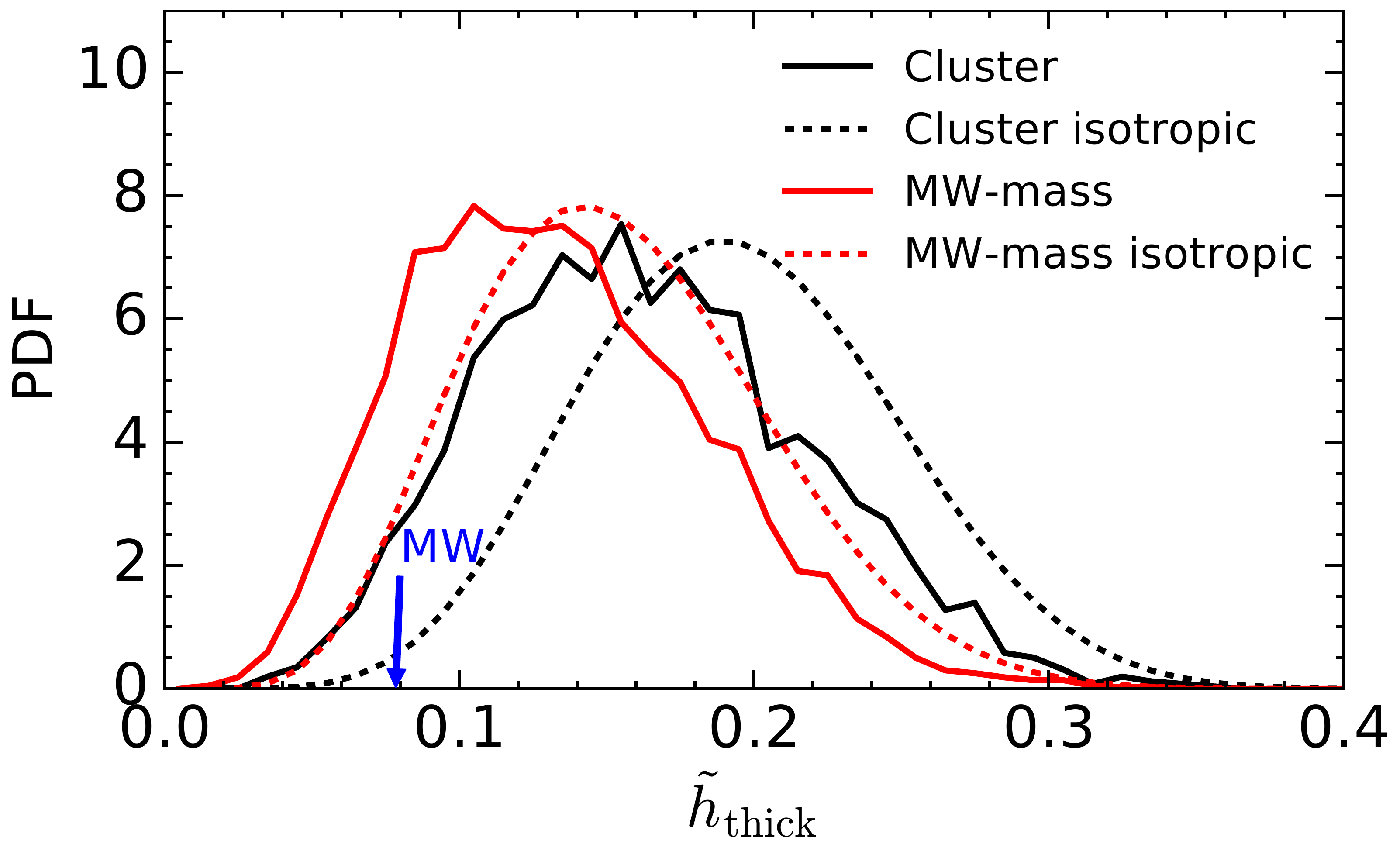}
\includegraphics[width=\columnwidth]{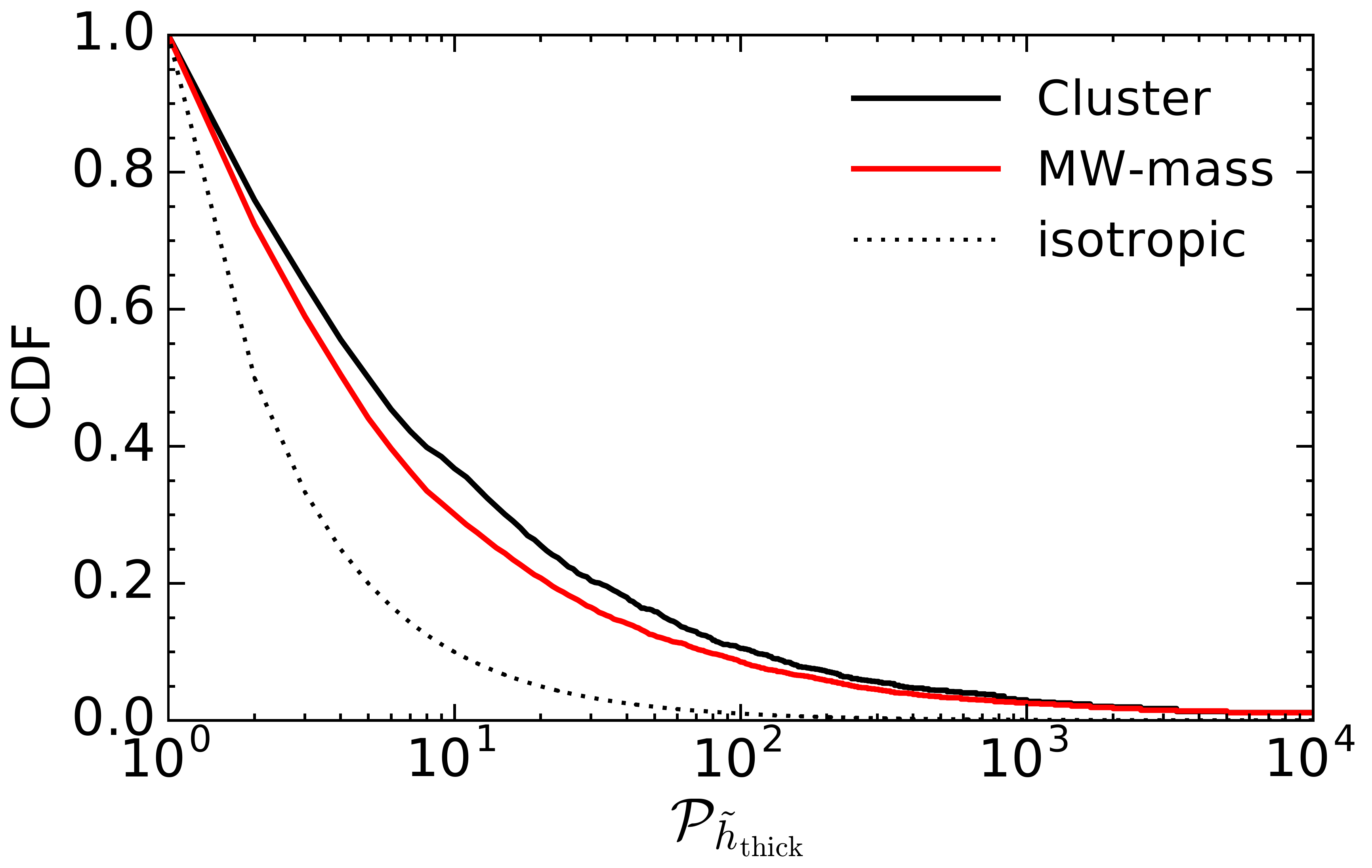}
\caption{\textit{Top panel}: the PDF of the fractional thickness of the subsystem of the 11 satellites with highest stellar mass. The black solid line shows the result for clusters in the MS7, while the black dashed line shows the result for corresponding isotropic distributions. The red solid and red dashed curves show the results for MW-mass systems and their isotropic samples in the MSIIsc7. The vertical blue arrow indicates the MW's value, $\tilde{h}_{\rm thick} =$ 0.0785. \textit{Bottom panel}: the complementary CDF of the prominence, ${\mathcal{P}_{\tilde{h}_{\rm thick}}}$, of the thickness of satellite distributions for galaxy clusters (black solid curve) and MW analogues (red solid curve).} 
\label{thickness}
\end{center}
\end{figure}

We use an alternative quantity, the fractional thickness, to describe the flattening of the distribution of the top 11 satellite galaxies, as described in Section \ref{sec:thickness}. In the top panel of Fig. \ref{thickness}, the black solid curve shows the distribution of fractional thickness of the plane of satellites in clusters. Note that each fractional thickness, $\tilde{h}_{\rm thick}$, has been scaled to the size of the host halo, which we take as 1 Mpc and 0.3 Mpc for clusters and MW analogues. There is a clear excess at the low fractional thickness end compared to the isotropic distribution (black dashed curve). The MW-mass systems have systematically lower $\tilde{h}_{\rm thick}$ values than clusters which is a manifestation of the former being more radially concentrated. When comparing to the MW, for which $\tilde{h}_{\rm thick}$ $=$ 0.0785 shown as the vertical blue arrow, we find 4.7 per cent of clusters and 13.1 per cent of MW-mass systems to have even lower thickness. In the bottom panel, the black solid curve shows that there are 44.6\% of satellite systems that are thinner than the 16th percentile of the $\tilde{h}_{\rm thick}$ distribution of isotropized systems for clusters. In the case of M31, \cite{Pawlowski2013} showed the thickness of the plane of 14 dwarf satellite galaxies is $\tilde{h}_{\rm thick}=0.0473\ \pm\ 0.0007$. As the results of $c/a$, the satellite galaxies in clusters deviate more from the isotropic distribution compared to those in MW analogues (the bottom panel). The relatively stronger deviation from isotropic distributions for satellite galaxies in clusters is reflected also by the higher fractions of satellite systems with $\tilde{h}_{\rm thick}$ less than the 16, 50 and 84 percentiles of the distribution of isotropic samples shown in the third row of Table. \ref{tab:fraction}.

\subsection{Relation to the host halo}
\label{sec:relate to halo}

\begin{figure}
\begin{center}
\includegraphics[width=\columnwidth]{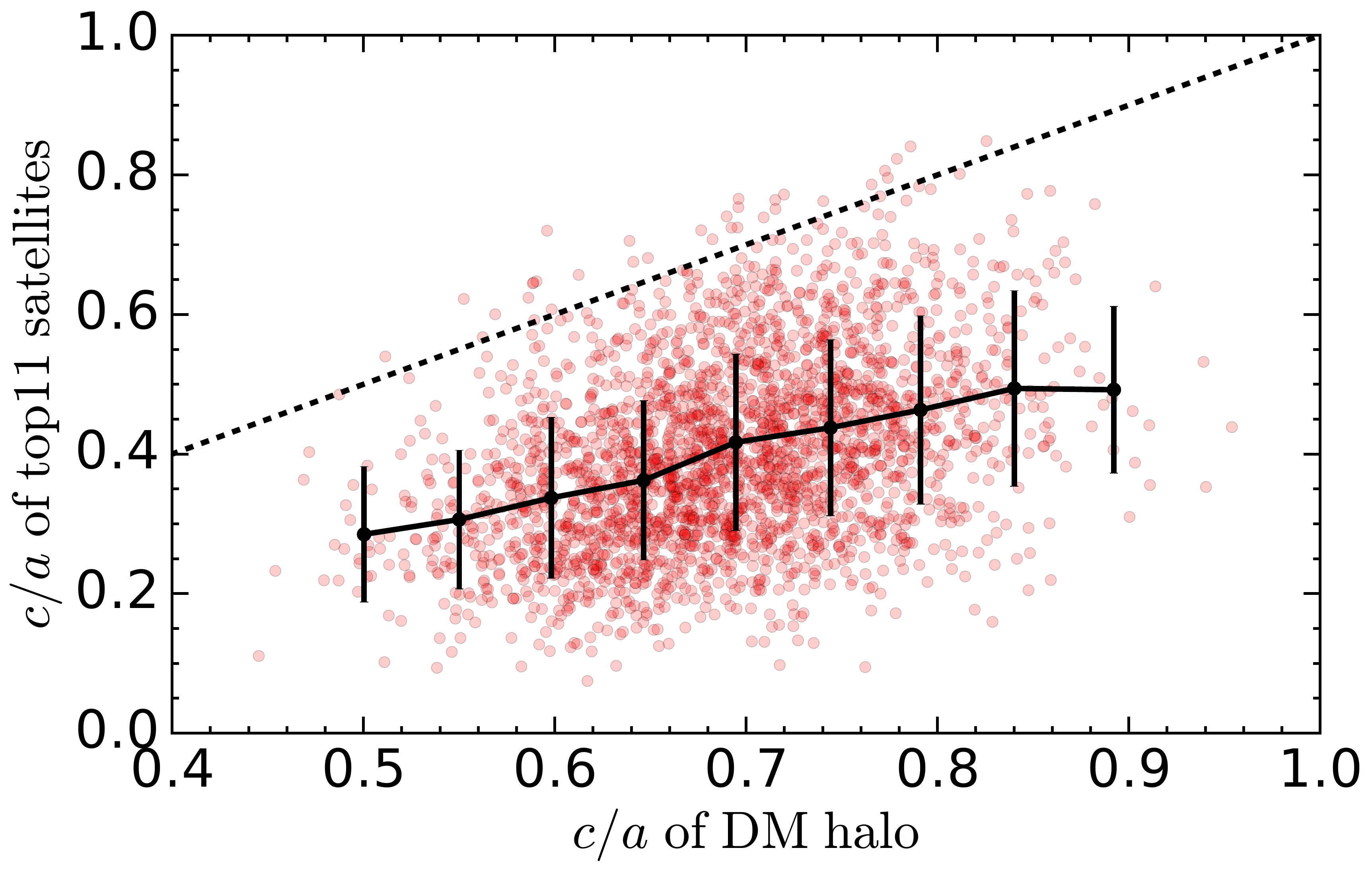}
\caption{Correlation between the axis ratios ($c/a$) of the subsystem of the 11 satellites with highest stellar mass with those of the host halo. The black solid curve and error bars represent the mean value and the 68 percentile scatter in the satellite $c/a$ as a function of the host $c/a$. The black dashed line shows the line of equality.}
\label{c/a-c/a}
\end{center}
\end{figure}

\begin{figure*}
\centering  
\includegraphics[width=0.45\textwidth]{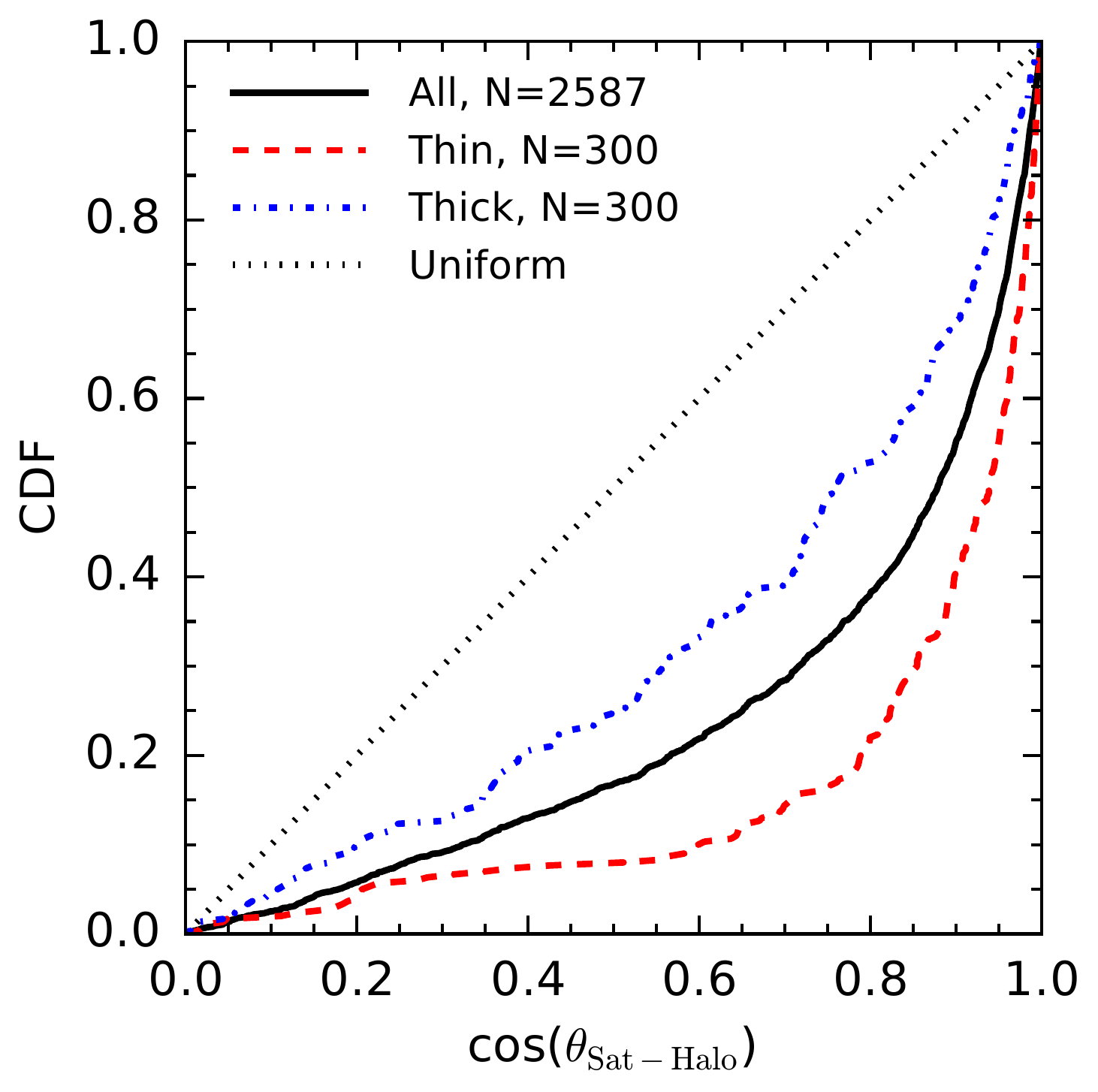}
\includegraphics[width=0.45\textwidth]{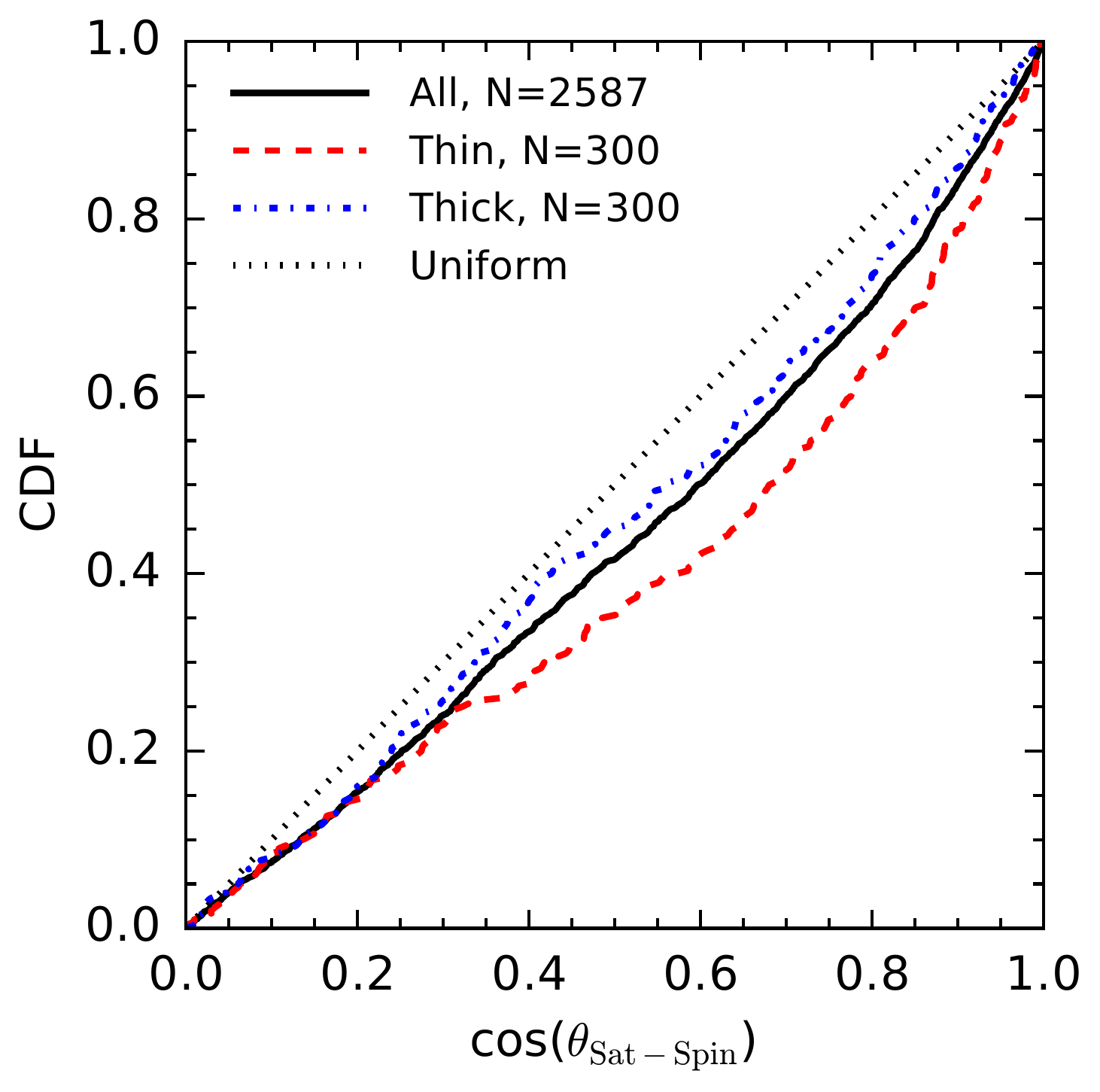}
\caption{\textit{Left panel}: the CDF of the misalignment angle, cos$\rm \theta_{Sat-Halo}$, between the plane of the subsystem of the 11 satellites with highest stellar mass and the orientation of the host halo. \textit{Right panel}: the CDF of the misalignment angle, cos$\rm \theta_{Sat-Spin}$, between the plane of the subsystem of the 11 satellites with highest stellar mass and the specific angular momentum of the host halo. The black dotted line shows the isotropic distribution. Results for the full sample, the thin-satellite-plane sample and the thick-satellite-plane sample are shown with the black solid curve, red dashed curve and blue dash-dotted curve, respectively.}
\label{fig_halo}
\end{figure*}

 In this section, we study the relation between the satellite distributions in clusters and their host halos. We first study the relation between the $c/a$ of the top 11 satellite configurations and the $c/a$ of their host halos. We then explore the relation between the direction of the plane of the top 11 satellites and the direction of the specific angular momentum and the orientation of the host halo. We refer to the normal vector of the plane of the top 11 satellites as its direction, and to the minor axis of the host halo as the halo's orientation. We use all DM particles within 1 Mpc from the centre of the halo to quantify the halo's shape and orientation (see Section \ref{sec:axis ratio}). We further divide our sample into three categories according to the fractional thickness of the plane of the top 11 satellites: full sample $-$ 2587 galaxy clusters; thin sample $-$ the thinnest 300 systems, which corresponds to $\tilde{h}_{\rm thick}$ < 0.10; and thick sample $-$ the thickest 300 systems (i,.e. $\tilde{h}_{\rm thick}$ > 0.23) and explore their relation to the properties of the DM halos.  

The relation between the $c/a$ of the DM halos and the $c/a$ of their top 11 satellite distributions is shown in Fig. \ref{c/a-c/a}. The mean values of $c/a$ and the corresponding standard deviations are shown by the black curve with error bars for a given halo $c/a$. We find a positive correlation between the halo shape and the shape of the top 11 satellites configuration, e.g. the flatter the halo, the flatter its satellite distribution. However, the scatter is rather large, about 30\% - 50\%. This finding suggests that satellite galaxies trace the DM in a rather stochastic manner.

We note that the situation could be more complex when taking into account baryonic processes. Previous works have shown that while the inner regions of halos are rounder in hydrodynamical simulations than in dark matter-only (DMO) simulations, the halo outskirts are largely unaffected \citep{Bryan2013,Butsky2016,Suto2017,Chua2019,Chua2021}. Moreover, the population of satellite galaxies is affected by enhanced tidal disruption due to the presence of a central galaxy which leads to a less concentrated radial distribution of satellites \citep{Garrison-Kimmel2017,Sawala2017,Richings2020}.
This results in higher $c/a$ ratios for the isotropic distributions since their radial density matches by construction that of the satellite galaxies. The confluence of these effects makes it difficult to estimate how the inclusion of realistic baryonic physics will affect the degree of anisotropy of the satellite distributions.

\citet{Knebe2004} studied the distribution of satellites in simulated clusters and found that the apocenters of the satellite orbits preferentially reside within a cone with an opening angle $\sim$ 40$^{\circ}$ around the major axis of the host halo. Previous works also found an elongated disc-like structure composed of satellite galaxies aligned with the major axis of the DM halo for a wide range of halo masses \citep{Libeskind2005,Wang2008,Lovell2011,Cautun2015a,Huang2016}.

The left panel of Fig. \ref{fig_halo} shows a strong correlation between the direction of the plane of the top 11 satellites and the orientation of the host halo. In half of the systems, the angle between the satellite plane normal and the host halo minor axis is smaller than $28\fdg5$. \citet{Shao2016} investigated the alignment of the satellite populations in MW analogues using the EAGLE simulation and found that half of their sample have misalignment angles smaller than $33\fdg8$. The planes of satellites are therefore somewhat more aligned with their host halos in clusters compared with those in MW analogues. For the thin sample, the angles between the planes of satellites and the host halos are even smaller with half of them being smaller than $20\fdg0$. The same is true for MW-mass hosts, the flattest satellite systems show the strongest alignment with the shape of their host \citep{Shao2020}.

The right panel of Fig. \ref{fig_halo} shows the correlation between the direction of the plane of the top 11 satellites and the direction of the angular momentum of their host halo. In contrast to the strong correlation seen with the halo orientations, the alignment with the halo' angular momentum is rather weak with a median value of the misalignment angle of $53\fdg4$ ($60^\circ$ for isotropy). However, the difference between this alignment and a uniform distribution is statistically significant with a $p$-value of the KS test of 2.22$\times10^{-12}$. When focusing on the thinnest plane sample, we only find a slightly stronger alignment signal. 

\subsection{Relation to the large-scale structure}
\label{sec:alignment with lss}

\begin{figure*}
\centering
\includegraphics[width=0.45\textwidth]{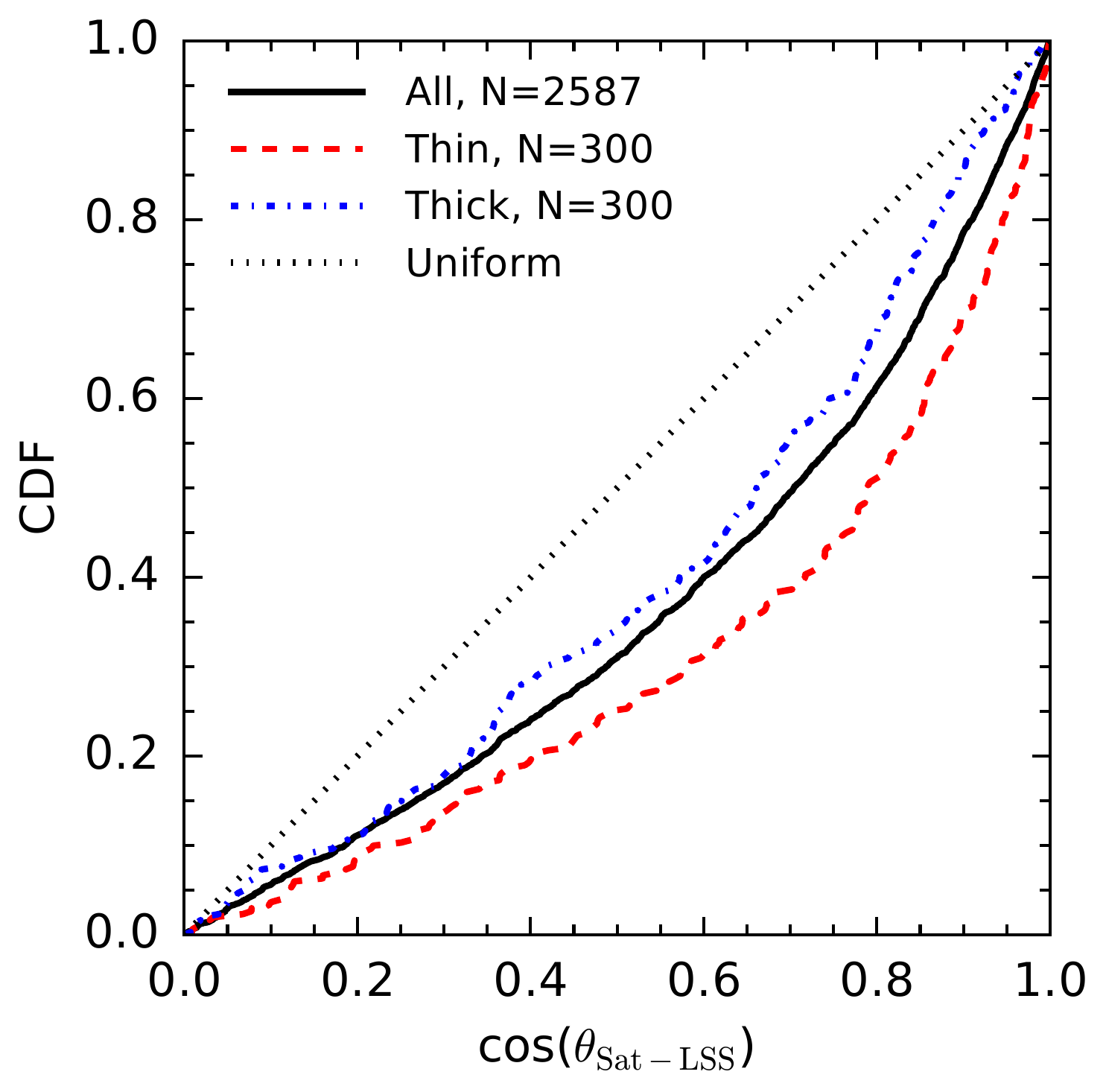}
\includegraphics[width=0.45\textwidth]{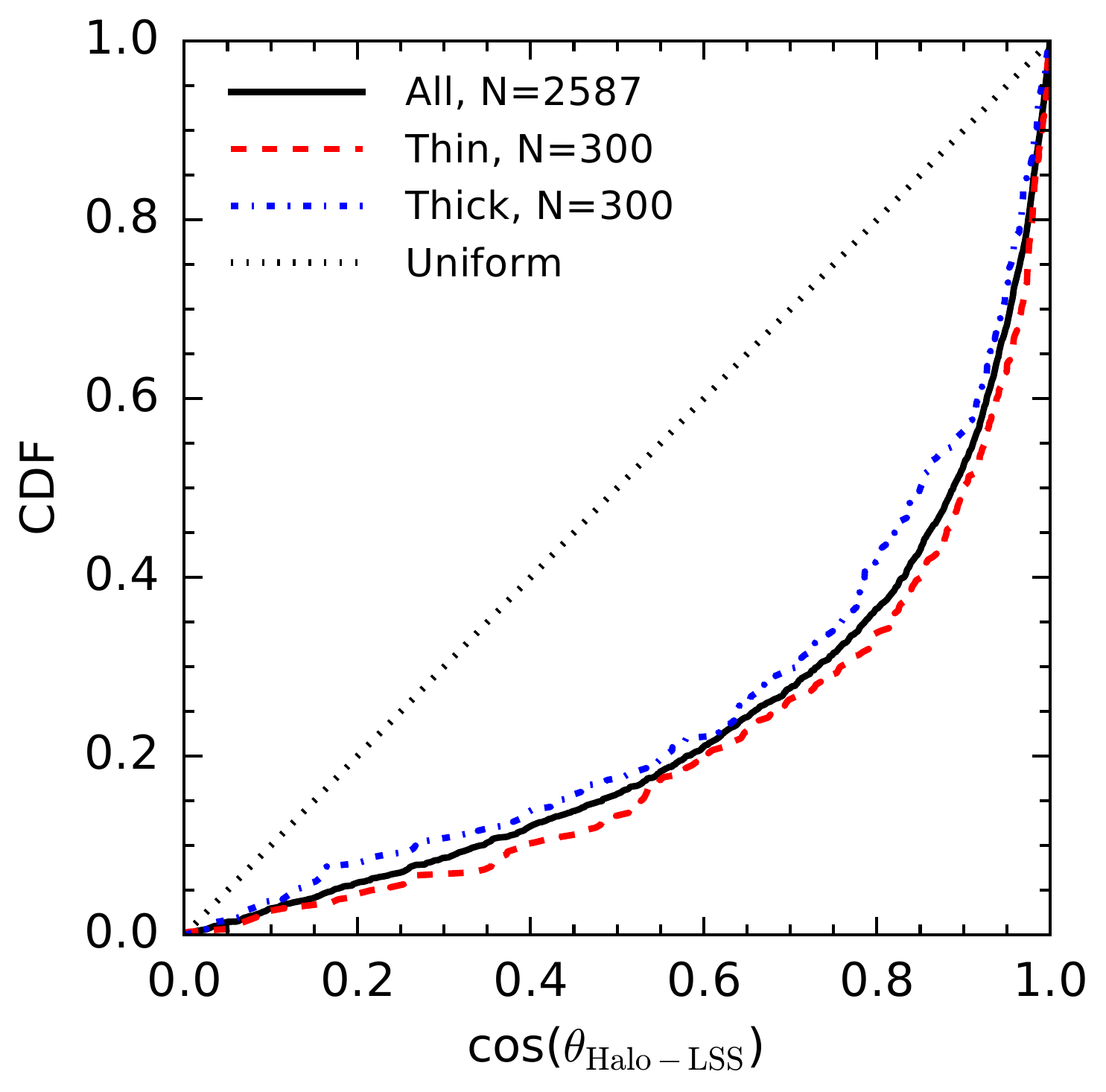}
\caption{\textit{Left panel}: the CDF of the misalignment angle, cos$\rm \theta_{Sat-LSS}$, between the plane of the subsystem of the 11 satellites with highest stellar mass and the surrounding LSS (on scales of 2$R_{\rm vir}$- 3$R_{\rm vir}$). \textit{Right panel}: the CDF of the misalignment angle, cos$\rm \theta_{Halo-LSS}$, between the host halo and LSS. The line styles and colours are as in Fig. \ref{fig_halo}.}
\label{fig_LSS}
\end{figure*}

The anisotropic distribution of satellite galaxies can arise from accretion along filaments \citep{Knebe2004,Zentner2005,Libeskind2005,2011,Lovell2011,Buck2015,Ahmed2017}. In this section, we investigate whether the filaments that preferentially feed the halo are related to the orientation of the plane of the top 11 satellites, as well as to the orientation of the host halo.

In Fig. \ref{fig_LSS}, it shows a mild alignment between the plane of the top 11 satellites and the LSS with a median value of the misalignment angle (60$^{\circ}$ for isotropy) of $45\fdg2$. This is somewhat smaller than the result for MW analogues for which the median value is $48\fdg7$ \citep{Shao2016}. The thin sample has a stronger alignment signal with a median angle of $37\fdg8$ than the thick sample with a median angle of $48\fdg3$. This suggests that a larger fraction of satellite galaxies may come in along filaments if the plane of satellites is thin. 

Previous works found the longest axis of the halo is aligned with the slowest collapsing eigenvector which is the same as the direction of the filament \citep[e.g.][]{Zhang2009, Wang2011, Libeskind2013a, Forero-Romero2014, Chen2016, Ganeshaiah2018, Okabe2020, Kuchner2020}. We find similar alignments as shown in the right panel of Fig. \ref{fig_LSS}. The median value of the misalignment angle between the halo orientation and the direction of the surrounding LSS is $27\fdg3$. The alignment signal is much stronger in clusters than in MW analogues which have a median value of $38\fdg7$ (as extracted from the work of \cite{Shao2016}, who used the same definition of the LSS). The alignment between the dark halo and the LSS is much stronger than the alignment between the plane of the top 11 satellites and the LSS, similar to what has been found for MW analogues \citep{Shao2016}. Given that the orientation of the plane of satellites is strongly aligned with the halo's direction (see the left panel of Fig. \ref{fig_halo}), the apparent alignment between the plane of satellites and the LSS can be caused by the strong correlation between the dark halo and the LSS. Interestingly, the dependence of the alignment level between the dark halo and the LSS on the thickness of the plane of satellites is very weak, potentially excluding the filamentary accretion as the main factor that determines the thickness of the plane of satellites.

\subsection{The spatial distribution of satellite galaxies at the time of accretion}
\label{sec:infall shape}

\begin{figure}
\begin{center}
\includegraphics[width=\columnwidth]{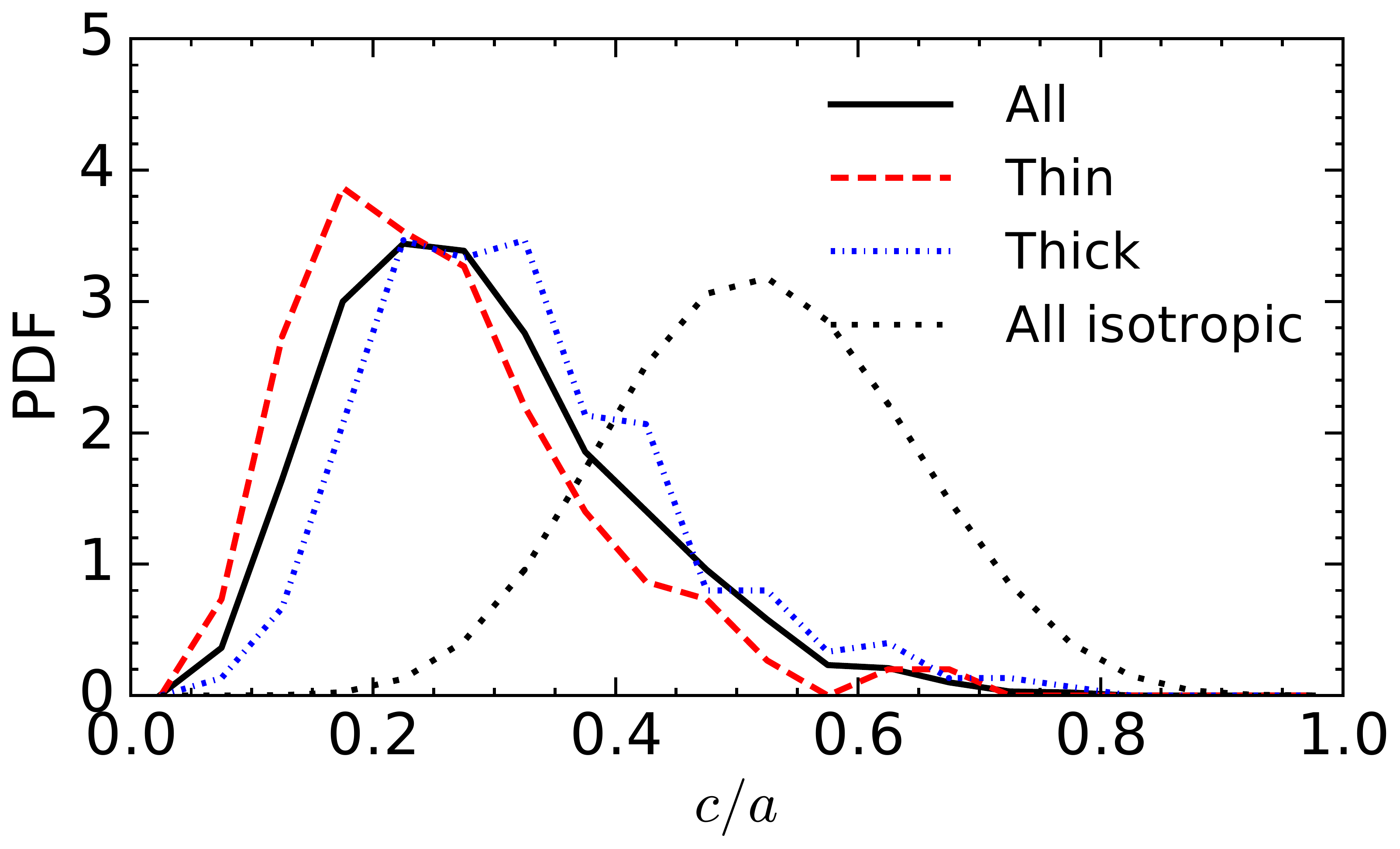}
\includegraphics[width=\columnwidth]{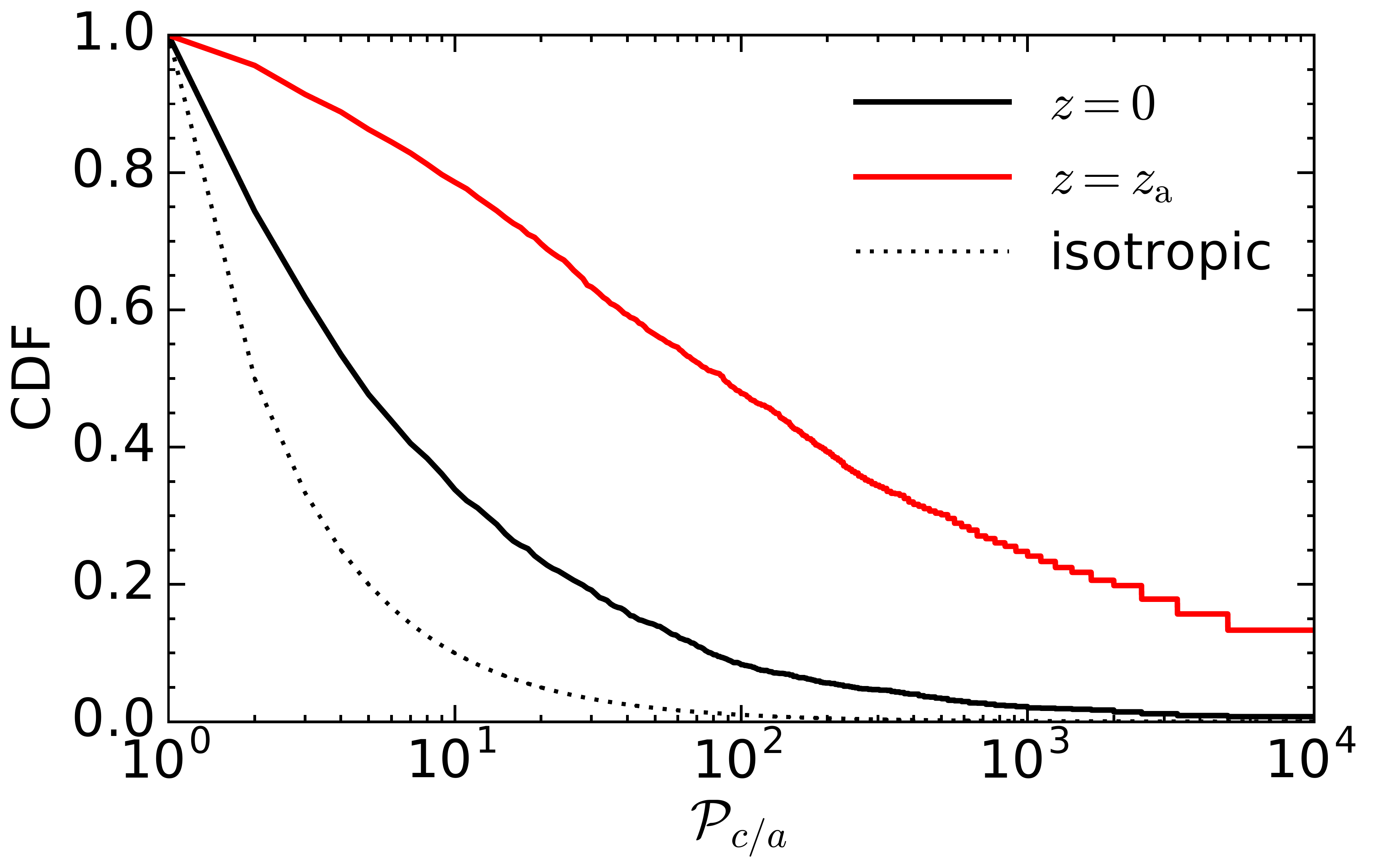}
\caption{\textit{Top panel}: the PDF of $c/a$ for the subsystem of the 11 satellites with highest stellar mass at their time of accretion. The black solid, red dashed, and blue dash-dotted lines are for: the full sample, the thin-satellite-plane sample and the thick-satellite-plane sample, respectively. The black dotted line shows the expectation for the full sample if the satellites were accreted isotropically. \textit{Bottom panel}: the complementary CDF of the prominence, $\mathcal{P}_{c/a}$, of the axis ratio. The black solid curve corresponds to the result at $z = 0$ and the red solid curve shows the result at the time of accretion.}
\label{c/a_infall}
\end{center}
\end{figure}

    A substantial fraction of matter is accreted onto clusters along filaments. In order to investigate whether the top 11 satellites are accreted along with special directions, we follow \citet{Shao2018} to measure the shape of the entry points of these satellites into their host halo. For each satellite and its central galaxy, starting from $z=0$ we trace their formation history using the MS7 merger trees until the snapshot where the satellite and central galaxy are not in the same FOF group. The entry point is defined as the position relative to the halo centre at this snapshot. The snapshot next to it is defined as the accretion snapshot and the corresponding redshift is defined as the accretion time $z_{\rm a}$. Sometimes a galaxy can have a large apocenter when orbiting a more massive system and thus can enter the host's radius multiple times. In this case, we define the accretion time as the last time it entered the cluster. We calculate the axis ratio, $c/a$, of the entry points using the same method as described in Section \ref{sec:axis ratio}. The result is shown in the top panel of Fig. \ref{c/a_infall}. We find that the entry points of satellite galaxies have very flat configurations compared to the isotropic accretion distributions. This is more clearly shown by the red solid curve in the bottom panel of Fig. \ref{c/a_infall}. We find that 84.2 per cent of the satellite systems have lower $c/a$ than the 16th percentile of the $c/a$ distribution for the isotropic case. The excess of large values of the prominence of $c/a$ of satellite configurations is much more significant at the time of accretion than at $z = 0$. 

We further compare the $c/a$ at accretion for the thin sample and the thick sample and find the dependence on thickness is only mild. The status at accretion alone does not explain the difference in the thickness of the plane of the top 11 satellites. One potential explanation is that the thickness of the plane of satellites varies rapidly with time since many of the satellites do not orbit within the plane of satellites \citep{Shao2016,Shao2019,Fernando2017}.

\subsection{Anisotropic distribution of satellite galaxies at high redshift}
\label{sec:shape at high z}

\begin{figure}
\centering 
\includegraphics[width=\columnwidth]{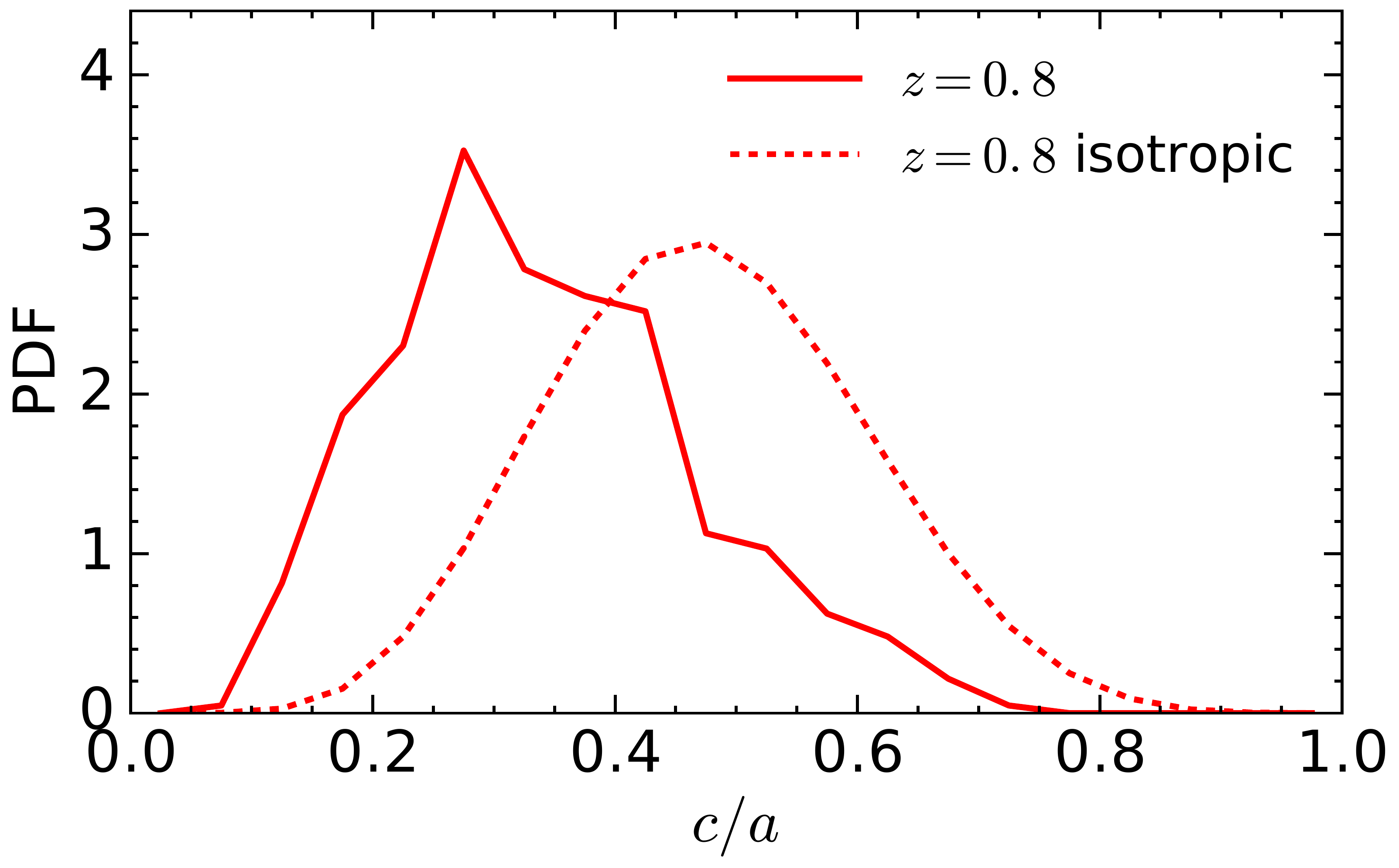}
\includegraphics[width=\columnwidth]{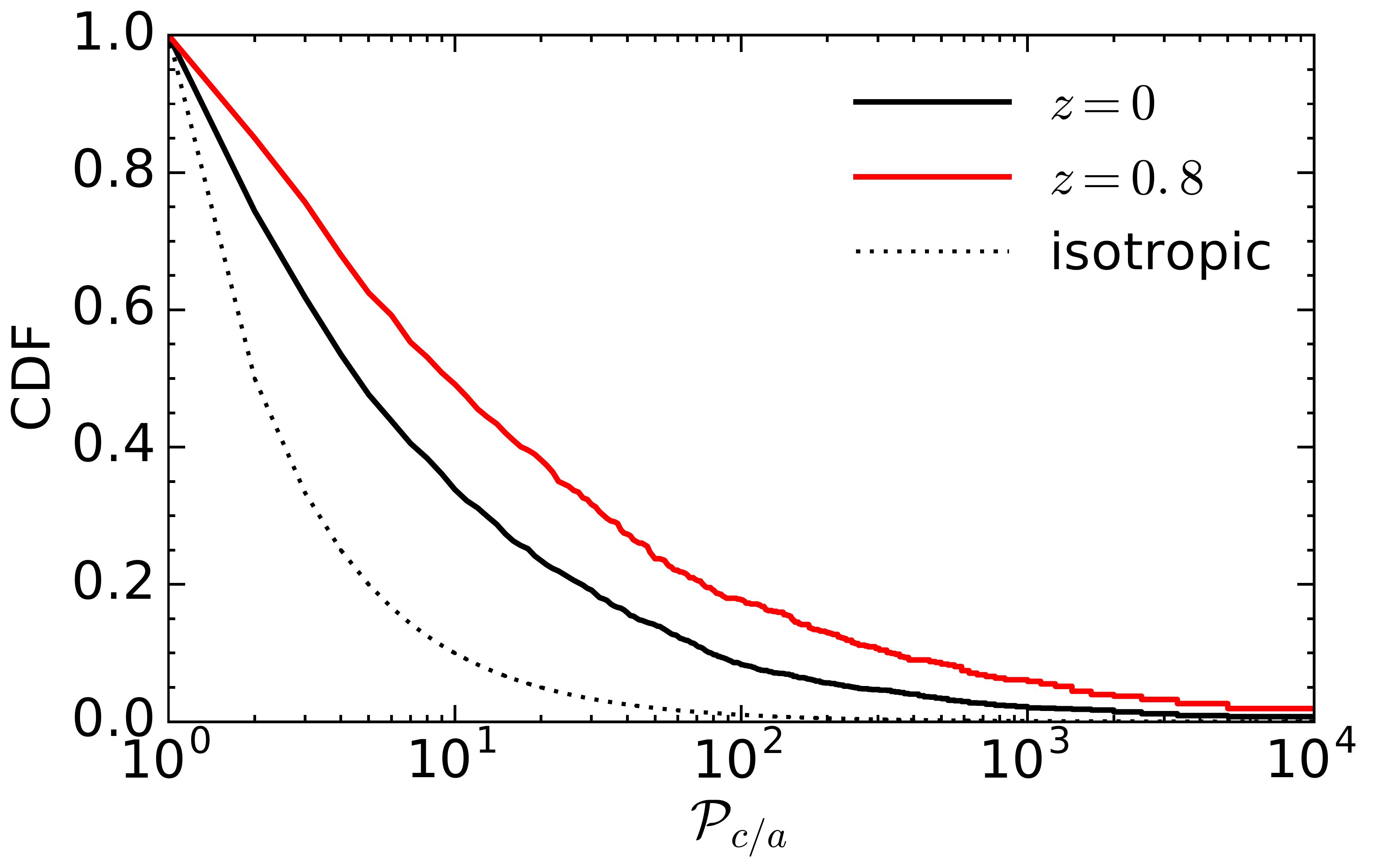}
\caption{\textit{Top panel}: the PDF of the axis ratio of the subsystem of the 11 satellites with highest stellar mass selected at $z =$ 0.8 (red solid line). The red dashed line shows the corresponding isotropic distribution. \textit{Bottom panel}: the complementary CDF of the prominence of the axis ratio at $z =$ 0 shown by the black solid line and at $z =$ 0.8 by the red solid line.}
\label{fig_c/a_z0.8}
\end{figure}

We saw that the top 11 satellite galaxies have a stronger anisotropy at accretion than their $z = 0$ descendants. Here we investigate how the anisotropy varies with redshift. We focus on satellite galaxies in galaxy clusters in the snapshot of 41 which corresponds to $z \sim 0.8$ and then for clarity we refer to it as $z = 0.8$. We select 834 galaxy clusters with halo mass $M_{\rm vir}\in{}\left(1,\ 3 \right)\times10^{14}$ $\rm M_{\odot}$ and have at least 11 satellite galaxies above $10^{9.5}$ $\rm M_{\odot}$ in that snapshot. At $z = 0.8$ the typical cluster size is smaller than at $z = 0$ and we thus use 812 kpc instead of 1 Mpc to select cluster members. The distributions of $c/a$ are shown in the top panel of Fig. \ref{fig_c/a_z0.8}. This shows a clear deviation from the isotropic distributions at high redshift. When compared to the results at $z = 0$, we find that the probability of finding a given prominence value of $c/a$ is higher at $z = 0.8$ than at $z = 0$ (see the bottom panel of Fig. \ref{fig_c/a_z0.8}). However, the level of anisotropy of the top 11 satellite galaxies in clusters selected at $z = 0.8$ is much lower than that at the accretion points of the top 11 satellite galaxies selected at $z = 0$, since these may have been accreted at a different time.

\subsection{The multiplicity of accreted groups}
\label{sec:group infall}

\begin{figure}
\centering
\includegraphics[width=\columnwidth]{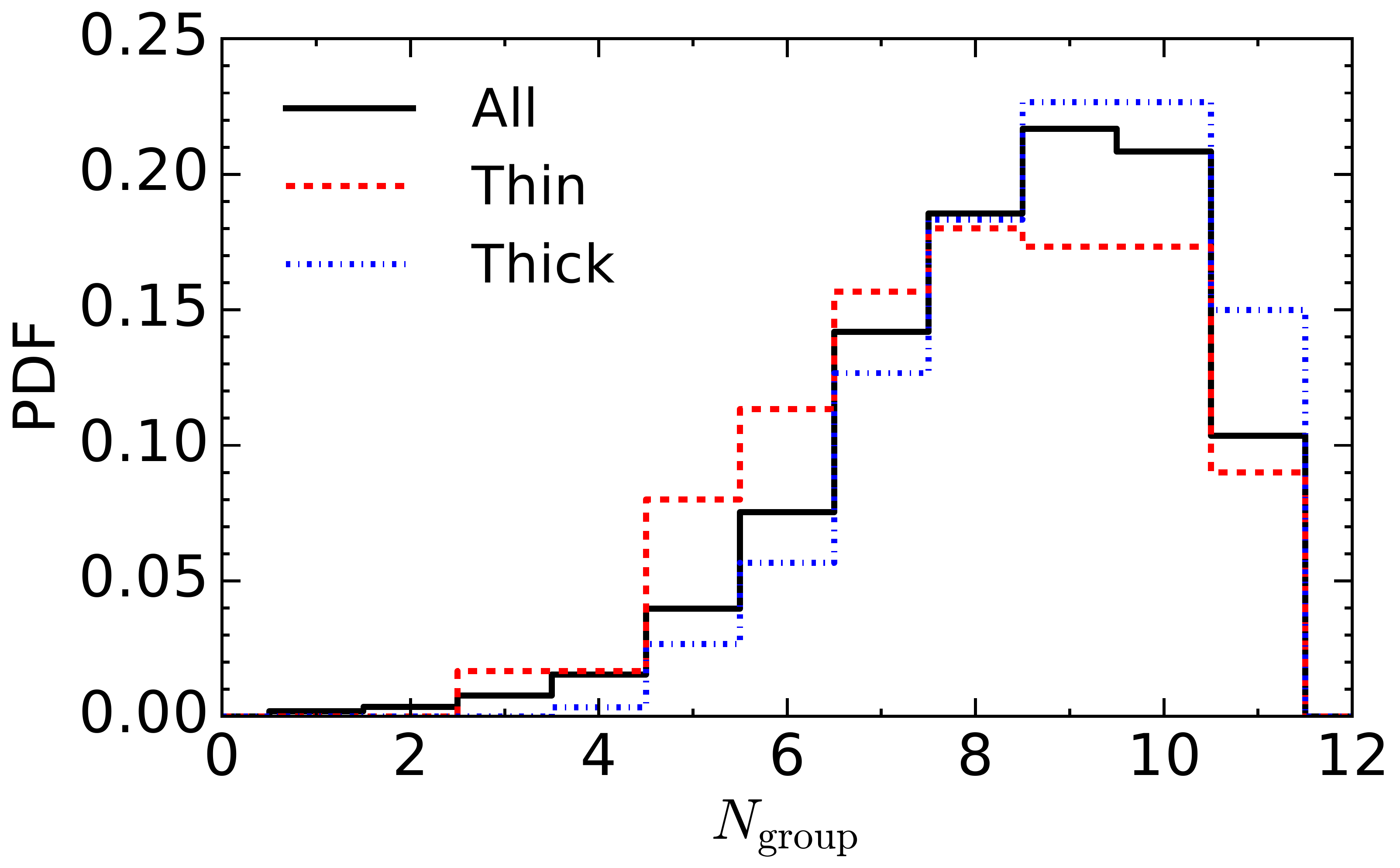}
\includegraphics[width=\columnwidth]{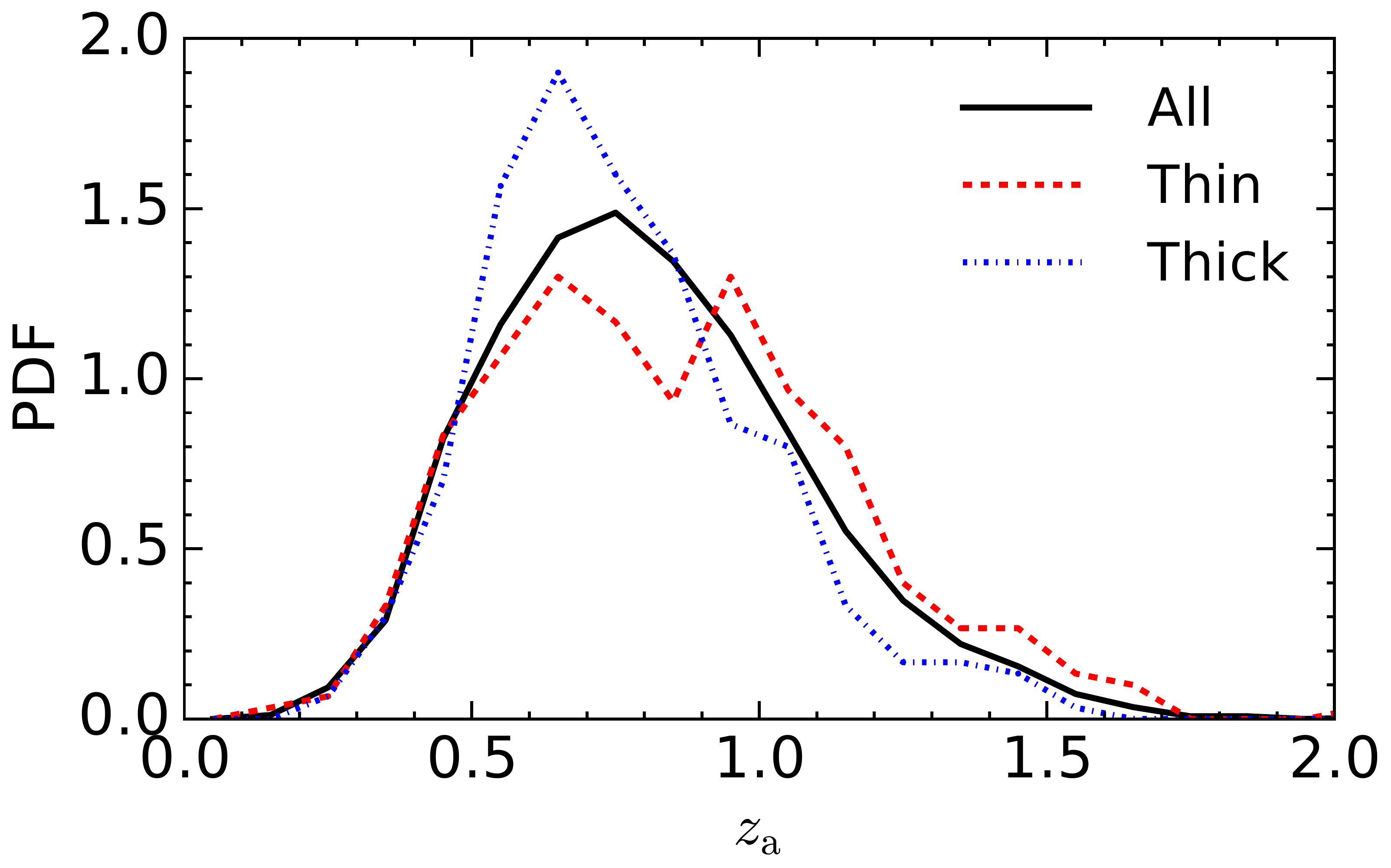}
\caption{\textit{Top panel}: the PDF of the number of separate groups at the time of accretion which originally hosted the 11 satellites with the highest stellar mass selected at $z = 0$. The black solid line corresponds to the full sample, the red dashed to the thin and the blue dash-dotted line to the thick sample. \textit{Bottom panel}: The mean redshift of accretion of the subsystem of the 11 satellites with highest stellar mass selected at $z = 0$.}
\label{fig_infall}
\end{figure}

\citet{Li2008} showed that satellite galaxies tend to reside in a flat plane if they  have been accreted in `groups' that share a similar infall time and orbital angular momentum. To estimate this effect, we trace the histories of the top 11 satellite galaxies up to the moment of infall. If multiple satellite galaxies belong to an existing group at the snapshot just before accretion onto the cluster, we regard their infall as a single group infall. We count the number of groups, $N_{\rm group}$, in which the top 11 satellites are distributed at the snapshot just before accretion and show their probability distribution in the top panel of Fig. \ref{fig_infall}. This shows that typically the top 11 satellite galaxies belong to 9 or 10 individual groups before being accreted into the cluster. For the thin sample, they are grouped in slightly fewer groups compared to the thick sample. There are very rare cases (0.2\%) when all of the 11 satellite galaxies belong to one particular group before being accreted onto the cluster. We conclude that most of the satellite galaxies in cluster-size halos are accreted one at a time and that group infall is not an important driver for the formation of planes of satellites. This is similar to the conclusions for satellite systems in MW-size halos \citep{Wang2013,Shao2018}.

One might expect that the thin-plane satellite galaxies were accreted later than the thick-plane satellite galaxies since in the former case the satellite galaxies might be more likely to preserve the information at infall. We test this hypothesis by studying the dependence of the thickness of the plane of satellites on the infall time in the bottom panel of Fig. \ref{fig_infall}. For each cluster, we calculate the mean redshift, $z_{\rm a}$, at the time of accretion (definition in Section \ref{sec:infall shape}) of the top 11 satellites which are selected at $z = 0$. Surprisingly, it shows that the probability distribution of the infall time of the thin sample is almost identical to that of the thick sample (although there is a small systematic difference), indicating that there is no correlation between the fractional thickness of the satellite galaxy plane and the infall time.

\section{Conclusions}
\label{sec:conclusion}

We use the Millennium simulation galaxy catalogue to investigate the spatial distribution of satellite galaxies in cluster-size halos, the relation with the properties of the host halo and the surrounding large-scale structure. For comparison, we also use the Millennium-II simulation galaxy catalogue to study MW-size halos. The two simulated galaxy catalogues are generated by implementing the \cite{Guo2013} semi-analytical galaxy formation model on the merger trees of the Millennium and Millennium-II simulations, which adopt \emph{WMAP7} cosmological parameters. There are 2587 cluster-size halos with $M_{\rm vir}\in{}\left(1,\ 3\right)\times10^{14}$ $\rm M_{\odot}$ with at least 11 satellite galaxies more massive than 3.16 $\times$ $10^{9}$ $\rm M_{\odot}$ and 4405 MW-size halos with $M_{\rm vir}\in{}\left(0.3,\ 3\right)\times10^{12}$ $\rm M_{\odot}$ with at least 11 satellite galaxies more massive than $10^{6}$ $\rm M_{\odot}$. We use the axis ratio and fractional thickness of the systems of the 11 most massive satellites to quantify the flattening of the satellite population. 

We use the MS7 mock catalogue to compare the projected distribution of satellites in clusters with the results of the SDSS. We select 516 galaxy clusters from the group catalogue of \cite{Yang2007}. We find that the MS7 is able to reproduce both the observed radial profiles traced by the 11 most massive satellite galaxies and the projected ellipticities of these top 11 satellite galaxies. Consistent with previous work, satellite galaxies show anisotropic distributions, with a typical ellipticity $\sim$ 0.2.

In simulations, the subsystems of the 11 most massive satellites are more anisotropic in clusters than in MW-mass hosts once these subsystems are compared to randomly isotropized versions of them. Nevertheless, values as extreme as $c/a = 0.183$ (the value for the MW's plane of satellites) are found in 3\% of the clusters. The difference between clusters and galactic halos is affected by cluster satellites being less radially concentrated than galactic satellites. We have accounted for this difference by calculating the prominence of the $c/a$ distribution with respect to the isotropic case with the exact same radial distribution of the top 11 satellite galaxies. On average, the satellite systems of clusters have higher prominence, i.e. are less likely to be the result of random fluctuations of an isotropic distribution, than those of MW-mass hosts.

Satellite distributions reflect their host DM halo in a complex and stochastic manner. The direction that is normal to the plane of satellites strongly aligns with the orientation of the host halo but only mildly with the halo spin. The planes are only weakly aligned with the local LSS, indicating that the correlation between planes of satellites and the surrounding filaments is non-trivial. The thinnest planes of satellites show consistently the largest alignment with the host halo shape and the LSS, although the difference with respect to the full population is rather small.

The distributions of satellite accretion points are very strongly anisotropic compared to their present distributions. However, this high degree of anisotropy is considerably reduced after accretion potentially due to differences in the orbital planes of different satellites, interactions with massive satellites, and torques from the host halo \citep{Bowden2013}. This means that satellite systems at high redshift, e.g at $z = 0.8$, are only mildly more anisotropic than in the present day.

We also investigate whether the plane of satellites may be caused by the accretion of satellites in a single group. We find that group infall cannot account for the satellite planar distributions as satellites are mostly accreted onto the halo individually rather than in groups, similar to what is found for MW analogues \citep{Wang2013,Shao2018}.

We have shown that in $\Lambda$CDM planes of satellites should be found not only in galactic halos but also in clusters of galaxies, which opens up the opportunity to study this topic more easily in a large sample of systems since clusters have brighter satellites that can be observed to larger distances and higher redshift than galactic halos.

\section*{Acknowledgements}

We thank the referee, Gary Allan Mamon, for his detailed and constructive comments which helped to improve the manuscript. This work is supported by the National Key Research and Development of China (grant number 2018YFA0404503) and the National Natural Science Foundation of China (NSFC; 12033008 and 11622325). CSF acknowledges support by the European Research Council (ERC) through 
Advanced Investigator grant DMIDAS (GA 786910) and also by the Science and 
Technology Facilities Council (STFC) through Consolidated Grant 
ST/P000541/1. This work used the DiRAC@Durham facility managed by the 
Institute for Computational Cosmology on behalf of the STFC DiRAC HPC 
Facility (www.dirac.ac.uk). The equipment was funded by BEIS capital 
funding via STFC capital grants ST/K00042X/1, ST/P002293/1, ST/R002371/1 
and ST/S002502/1, Durham University and STFC operations grant 
ST/R000832/1. DiRAC is part of the National e-Infrastructure.
CGL acknowledges support from STFC
(ST/T000244/1). MC is supported by the EU Horizon 2020 research and innovation programme under a Marie Sk{\l}odowska-Curie grant agreement 794474 (DancingGalaxies).
%%%%%%%%%%%%%%%%%%%%%%%%%%%%%%%%%%%%%%%%%%%%%%%%%%

\section*{DATA AVAILABILITY}

The data produced in this paper are available upon reasonable request to the corresponding author.


\bibliographystyle{mnras}
\bibliography{paper} % if your bibtex file is called example.bib

%%%%%%%%%%%%%%%%%%%%%%%%%%%%%%%%%%%%%%%%%%%%%%%%%%

%%%%%%%%%%%%%%%%% APPENDICES %%%%%%%%%%%%%%%%%%%%%

\appendix
\section{SATELLITE AXIS RATIO AS A FUNCTION OF THICKNESS}

\begin{figure}
\begin{center}
\includegraphics[width=\columnwidth]{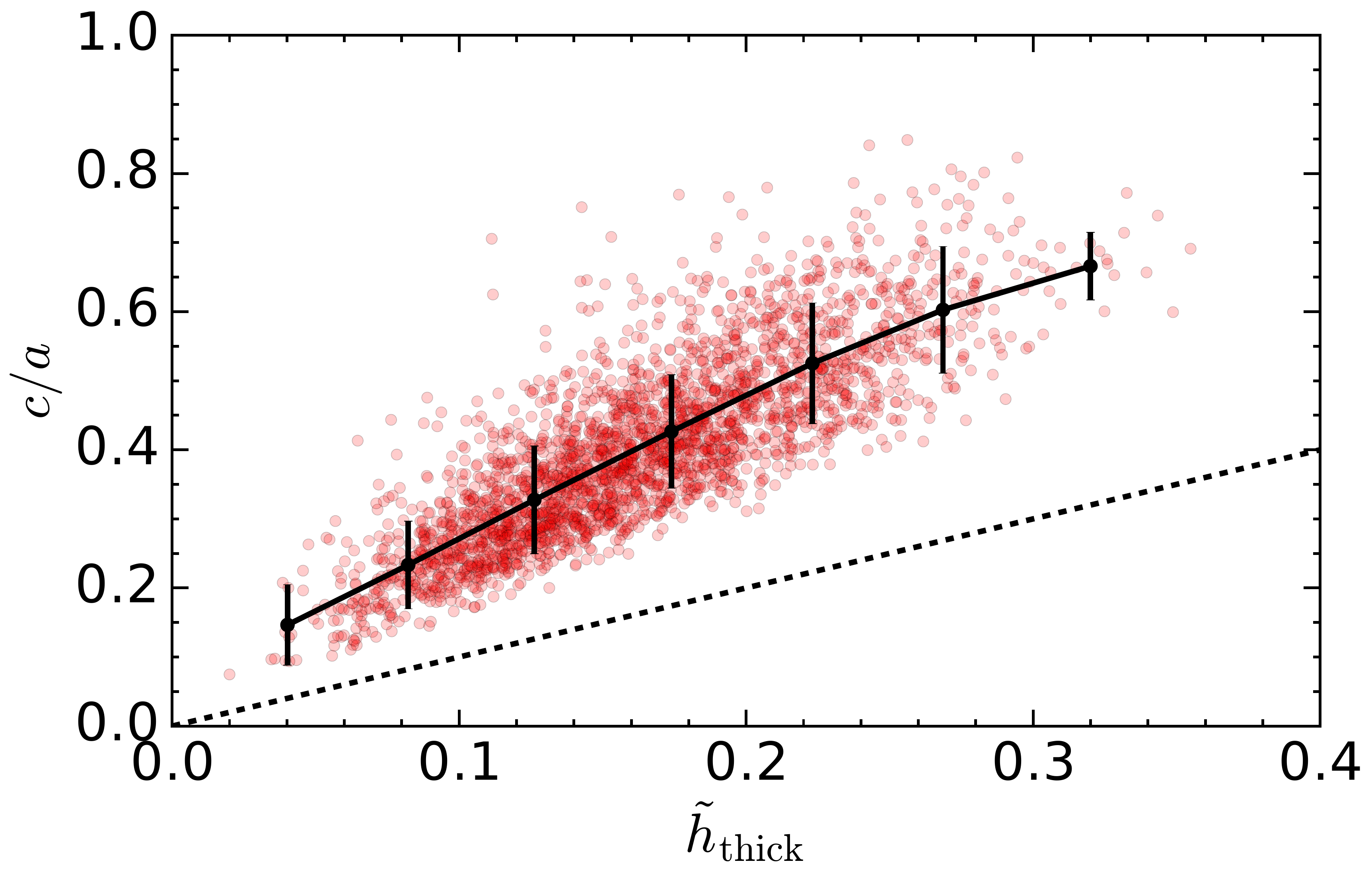}
\caption{Correlation between the axis ratio, $c/a$, and the thickness, $\tilde{h}_{\rm thick}$, of the subpopulation of 11 satellites with the highest stellar mass. The black solid curve and error bars represent the mean value and the 68 percentile scatter in the satellite $c/a$ as a function of the $\tilde{h}_{\rm thick}$. The black dashed line shows the line of equality.}
\label{c/a-h}
\end{center}
\end{figure}

In this work, we have used both the axis ratio and thickness to quantify the shape of satellite distributions. These two parameters are not independent, and indeed $\tilde{h}_{\rm thick}$ is the same as the short axis $c$ up to a constant scaling factor. We compare $c/a$ and $\tilde{h}_{\rm thick}$ in Fig. \ref{c/a-h}, where the black solid curve and error bars show the mean and the 68 percentile scatter in bins of the $\tilde{h}_{\rm thick}$ parameter. We find a considerable correlation with a moderate scatter between $c/a$ and $\tilde{h}_{\rm thick}$.

%If you want to present additional material which would interrupt the flow of the main paper, it can be placed in an Appendix which appears after the list of references.

%%%%%%%%%%%%%%%%%%%%%%%%%%%%%%%%%%%%%%%%%%%%%%%%%%

% Don't change these lines
\bsp	% typesetting comment
\label{lastpage}
\end{document}